  \documentclass[twocolumn,showpacs,aps,epsfig,nofootinbib]{revtex4}

\usepackage{graphicx}
\usepackage{epstopdf}
\usepackage{latexsym}
\usepackage{amssymb}
\usepackage{braket}
\usepackage{color}

\usepackage[center]{subfigure}

\begin{document}

%%%%%%%%%%%%%%%%%%%%%%%%%%%%%%%%%%%%%%%%%%%%%%%%%%%%%%%%%%%%%%%
 \newcommand{\bq}{\begin{equation}}
 \newcommand{\eq}{\end{equation}}
 \newcommand{\bqn}{\begin{eqnarray}}
 \newcommand{\eqn}{\end{eqnarray}}
 \newcommand{\nb}{\nonumber}
 \newcommand{\lb}{\label}
\newcommand{\PRL}{Phys. Rev. Lett.}
\newcommand{\PL}{Phys. Lett.}
\newcommand{\PR}{Phys. Rev.}
\newcommand{\CQG}{Class. Quantum Grav.}
 %%%%%%%%%%%%%%%%%%%%%%%%%%%%%%%%%%%%%%%%%%%%%%%%%%%%%%%%%%%%%%%

\title{Inflation in  general covariant theory of gravity}    

\author{Yongqing Huang ${} ^{a}$}
\email{yongqing_huang@baylor.edu}

\author{Anzhong Wang ${} ^{a,b}$}
\email{anzhong_wang@baylor.edu}

 \author{Qiang Wu   ${} ^{b}$}
\email{wuq@zjut.edu.cn}

\affiliation{ ${} ^{a}$GCAP-CASPER, Physics Department, Baylor
University, Waco, TX 76798-7316, USA\\  
${} ^{b}$ Institute for Advanced Physics $\&$ Mathematics, Zhejiang University of
Technology, Hangzhou 310032,  China  }

\date{\today}

\begin{abstract}

In this paper, we study inflation  in the framework of the nonrelativistic general covariant theory of the Ho\v{r}ava-Lifshitz gravity 
with the projectability condition and  an arbitrary coupling constant $\lambda$. We find that  the Friedmann-Robterson-Walker 
(FRW) universe is necessarily flat in such a setup. We work out explicitly the linear perturbations of the flat FRW universe without 
specifying to a particular gauge, and find that the perturbations are different from those obtained in general relativity, because of 
the presence of the high-order spatial derivative terms. Applied the general formulas to a single scalar field, we show that in the 
sub-horizon regions, the metric and scalar field are tightly coupled and have the same oscillating frequencies. In the super-horizon 
regions,  the perturbations become adiabatic, and the comoving curvature perturbation is constant. We also calculate the power 
spectra and  indices of both the scalar and tensor perturbations, and  express them explicitly in terms of the slow roll parameters 
and the coupling constants of the high-order spatial derivative terms. In particular, we find that the perturbations, of both scalar  
and tensor, are almost scale-invariant, and, with some reasonable assumptions on the coupling coefficients,  the 
spectrum index of the tensor perturbation is the same as that given in the minimum scenario in GR, whereas the index for scalar 
perturbation in general depends on $\lambda$ and is different from the standard GR value.  The ratio of the scalar and 
tensor power spectra depends on the  high-order spatial derivative terms, and can be different from that of GR significantly.

\end{abstract}

%; 98.80.Cq} %04.60.-m;} 98.80.-k;

\maketitle

\section{Introduction}
\renewcommand{\theequation}{1.\arabic{equation}} \setcounter{equation}{0}

The   Ho\v{r}ava-Lifshitz (HL) theory of   quantum gravity, proposed recently by Ho\v{r}ava \cite{Horava}, motivated by the Lifshitz 
scalar field theory in solid state physics \cite{Lifshitz}, has attracted a great deal of attention, due to
its several remarkable features \cite{reviews,Mukc}. 
 The  HL  theory  is based on the perspective that Lorentz symmetry should appear as an emergent symmetry at long 
distances, but can be fundamentally  absent at short ones \cite{Pav}. In the latter regime,  the system 
 exhibits a strong anisotropic scaling between space and time, 
\bq
\lb{1.1}
{\bf x} \rightarrow \ell {\bf x}, \;\;\;  t \rightarrow \ell^{z} t,
\eq
where $z \ge 3$ in the $(3+1)$-dimensional spacetime \cite{Horava,Visser}. At long distances, high-order curvature corrections become
negligible, and  the lowest order terms   $R$ and $\Lambda$ take over,  whereby the Lorentz invariance is expected to be
 ``accidentally restored," where $R$ denotes the 3-dimensional Ricci scalar of the hypersurfaces $t =$ Constant, and $\Lambda$ the cosmological constant.
 
Because of the anisotropic scaling,   the gauge symmetry of the theory is broken down to the  foliation-preserving 
diffeomorphism, Diff($M, \; {\cal{F}}$),  
\bq
\lb{1.4}
\delta{t} =  - f(t),\; \;\; \delta{x}^{i}  =    - \zeta^{i}(t, {\bf x}),
\eq
for which the lapse function $N$, shift vector  $N^{i}$, and 3-spatial metric $g_{ij}$   transform as
\bqn
\lb{1.5}
\delta{N} &=& \zeta^{k}\nabla_{k}N + \dot{N}f + N\dot{f},\nb\\
\delta{N}_{i} &=& N_{k}\nabla_{i}\zeta^{k} + \zeta^{k}\nabla_{k}N_{i}  + g_{ik}\dot{\zeta}^{k}
+ \dot{N}_{i}f + N_{i}\dot{f}, \nb\\
\delta{g}_{ij} &=& \nabla_{i}\zeta_{j} + \nabla_{j}\zeta_{i} + f\dot{g}_{ij}, 
\eqn
where $\dot{f} \equiv df/dt,\;  \nabla_{i}$ denotes the covariant 
derivative with respect to   $g_{ij}$,  $N_{i} = g_{ik}N^{k}$, and $\delta{g}_{ij} 
\equiv \tilde{g}_{ij}\left(t, x^k\right) - {g}_{ij}\left(t, x^k\right)$,
 etc. From these expressions one can see that   $N$ and   $N^{i}$ play the role of gauge fields of the Diff($M, \; {\cal{F}}$). 
 Therefore, it is natural to assume that $N$ and $N^{i}$ inherit the same dependence on 
space and time as the corresponding generators \cite{Horava}, 
\bq
\lb{1.6}
N = N(t), \;\;\; N^{i} = N^{i}(t, x),
\eq
which is  often referred to as the projectability condition.

Due to the  Diff($M, \; {\cal{F}}$) diffeomorphisms (\ref{1.4}), one more degree of freedom appears
in the gravitational sector - a spin-0 graviton. This is potentially dangerous, and needs to decouple  
in the IR regime, in order to be consistent with observations.   Whether this is possible or not is still an
open question \cite{reviews,PHH}.  In particular,  spherically symmetric static spacetimes were studied  in \cite{Mukc}, and shown
that the spin-0 graviton indeed decouples after nonlinear effects are taken into account, an analogue of the 
Vainshtein effect in massive gravity  \cite{Vain}. Along the same direction,  considerations in cosmology were  given  
in \cite{WWa,Izumi:2011eh,GMW}. In particular, in \cite{Izumi:2011eh,GMW} a fully nonlinear analysis of superhorizon cosmological perturbations was carried out,
by adopting the so-called gradient expansion method \cite{SBLMS}. It was found that   the relativistic  limit   is continuous,
and  general relativity (GR) is recovered in two different cases: (a) when only  the ``dark matter as an integration constant'' is present \cite{Izumi:2011eh};
and (b) when a scalar field and    the ``dark matter as an integration constant'' are present \cite{GMW}. 

Another very promising approach is to eliminate the spin-0 graviton by introducing two auxiliary fields,
the $U(1)$ gauge field $A$ and the Newtonian prepotentail $\varphi$, by   extending 
the  Diff($M, \; {\cal{F}}$) symmetry (\ref{1.4}) to include  a local $U(1)$ symmetry \cite{HMT}, 
\bq
\lb{symmetry}
 U(1) \ltimes {\mbox{Diff}}(M, \; {\cal{F}}).
 \eq
Under this extended symmetry,   the special status of time  maintains,  so that the anisotropic scaling (\ref{1.1})
can  still be  realized,  and the theory is kept power-counting renormalizable. Meanwhile, because of the elimination of the spin-0 graviton,  its IR  behavior can be 
 significantly improved.  Under the Diff($M, \; {\cal{F}}$), $A$ and $\varphi$ transform as,
\bqn
\lb{2.2}
\delta{A} &=& \zeta^{i}\partial_{i}A + \dot{f}A  + f\dot{A},\nb\\
\delta \varphi &=&  f \dot{\varphi} + \zeta^{i}\partial_{i}\varphi.
\eqn
Under the local $U(1)$ symmetry,  the fields  
 transform as
\bqn
\lb{2.3}
\delta_{\alpha}A &=&\dot{\alpha} - N^{i}\nabla_{i}\alpha,\;\;\;
\delta_{\alpha}\varphi = - \alpha,\nb\\ 
\delta_{\alpha}N_{i} &=& N\nabla_{i}\alpha,\;\;\;
\delta_{\alpha}g_{ij} = 0,\;\;\; \delta_{\alpha}{N} = 0,
\eqn
where $\alpha$ is   the generator  of the local $U(1)$ gauge symmetry. For the detail, we refer readers to \cite{HMT,WW}. 

The elimination of the spin-0 graviton was done  initially in the   case $\lambda = 1$ \cite{HMT,WW}, but soon generalized 
to the case with any $\lambda$ \cite{Silva,HW,LWWZ}, where $\lambda$ denotes a coupling constant that 
characterizes the deviation of  the kinetic part of action from the corresponding one given in GR with $\lambda_{GR} =1$
 (For the analysis of Hamiltonian consistency, see \cite{HMT,Kluson}).  To avoid the strong coupling problem, one may follow Blas,
 Pujolas and Sibiryakov (BPS) \cite{BPSc} to introduce an energy scale $M_{*}$ that satisfies the condition 
  \cite{LWWZ},
 \bq
 \lb{SC}
 M_{*} < \Lambda_{\omega},
 \eq
 where $M_{*}$ is the suppression energy of the sixth-order derivative terms, and $\Lambda_{\omega}$ is the would-be strong coupling energy
 scale, given by,
 \bq
 \lb{SCa}
 \Lambda_{\omega} \simeq \left(\frac{\zeta}{c_{1}}\right)^{3/2}|\lambda -1|^{5/4} M_{\text{pl}},
 \eq
 where   $\zeta$ is related to the Planck mass $M_{\text{pl}}$   through Eq.(\ref{NC}), and $c_{1}$, defined in (\ref{5.5}),  represents the coupling of a scalar
 field with the gauge field $A$. In the case without the projectability condition, the observed alignment of the rotation axis of the Sun with the ecliptic requires
 $M_{*} \lesssim 10^{15}\; {\mbox{GeV}}$ \cite{BPSc}. Similar considerations have not been carried out in the current version of the HL theory, and the up bound of 
 $M_{*}$ is unknown. From the above expression, it is clear that $\lambda$ cannot be precisely equal to one, in order for the BPS mechanism to work
 either without \cite{BPSc}  or with \cite{LWWZ}  the projectability condition.

 It is remarkable to note that the elimination of the spin-0 graviton can be also realized in the non-projectability 
 case with the extended  symmetry (\ref{symmetry}) \cite{ZWWS,ZSWW}. In addition, the number of independent coupling constants
 can be significantly reduced (from more than 70 \cite{KP} to 15), by simply imposing  the   softly breaking detailed balance condition,  while  
the theory  still remains power-counting renornalizable and has  a healthy IR limit.

 In this paper, we study inflation  of a scalar field in the Ho\v{r}ava and Melby-Thompson (HMT) setup with the projectability condition  \cite{HMT}  and 
 an arbitrary coupling constant $\lambda$  \cite{Silva}. Specifically, after  a brief review of the theory in Sec. II, we first show that the FRW
 universe is necessarily flat, when  it  is filled with (multi-) scalar, vector or fermionic fields in Sec. III.A.  
 Then, in the second part of Sec. III we present the general linear scalar perturbations without specifying to a particular gauge or specific  matter fields, while in the
 third part of it, we consider several possible gauge choices. Unlike the case without the U(1) symmetry \cite{WM}, some gauges used  in GR
 \cite{MW09},  such as the longitudinal gauge, now become possible, because of the U(1) gauge freedom. In Sec. IV, we first consider  
  the flat FRW background, and show clearly that the slow-roll conditions  imposed in GR are also
 needed here, in order to obtain enough e-fold to solve the problems such as horizon, monopole, domain walls, and so on \cite{Inflation}.  In addition, in this section we also show that
 in the super-horizon regions,  the perturbations become adiabatic, and the comoving curvature perturbation is constant. In Sec. V, we  
 show explicitly that in the sub-horizon regions, the metric and scalar field are tightly coupled and have the same oscillating frequencies, while in the 
super-horizon regions,  the perturbations are almost scale-invariant.  It is remarkable that a master
 equation for the scalar perturbations exists, in contrast to the case without the U(1) symmetry \cite{WM}. In Sec. VI, we calculate the power spectra and 
 indices of both scalar and tensor perturbations in the slow-roll approximations, by using the uniform    approximation  \cite{Hab}.
We    express  them explicitly in terms of the slow roll parameters and the coupling constants of the high-order spatial derivative terms.  
 Finally, in Sec. VII we present our main conclusions.

\section{General covariant theory with an arbitrary  constant $\lambda$}

\renewcommand{\theequation}{2.\arabic{equation}} \setcounter{equation}{0}

 In this section, we shall give a very brief introduction to the HMT setup with the projectability condition and $\lambda \not= 1$.
 For detail, we refer readers to   \cite{HMT,Silva,HW}.
 
 The total action of the theory can be written as,
 \bqn \lb{2.4}
S &=& \zeta^2\int dt d^{3}x N \sqrt{g} \Big({\cal{L}}_{K} -
{\cal{L}}_{{V}} +  {\cal{L}}_{{\varphi}} +  {\cal{L}}_{{A}} +  {\cal{L}}_{{\lambda}} \nb\\
& & ~~~~~~~~~~~~~~~~~~~~~~ \left. +{\zeta^{-2}} {\cal{L}}_{M} \right),
 \eqn
where $g={\rm det}(g_{ij})$, and
 \bqn \lb{2.5}
{\cal{L}}_{K} &=& K_{ij}K^{ij} -   \lambda K^{2},\nb\\
{\cal{L}}_{\varphi} &=&\varphi {\cal{G}}^{ij} \Big(2K_{ij} + \nabla_{i}\nabla_{j}\varphi\Big),\nb\\
{\cal{L}}_{A} &=&\frac{A}{N}\Big(2\Lambda_{g} - R\Big),\nb\\
{\cal{L}}_{\lambda} &=& \big(1-\lambda\big)\Big[\big(\Delta\varphi\big)^{2} + 2 K \Delta\varphi\Big].
 \eqn
Here $\Delta \equiv g^{ij}\nabla_{i}\nabla_{j}$,  
$\Lambda_{g}$ is a    coupling constant,  the
Ricci and Riemann tensors $R_{ij}$ and $R^{i}_{jkl}$  all refer to the 3-metric $g_{ij}$, and 
 \bqn \lb{2.6}
K_{ij} &=& \frac{1}{2N}\left(- \dot{g}_{ij} + \nabla_{i}N_{j} +
\nabla_{j}N_{i}\right),\nb\\
{\cal{G}}_{ij} &=& R_{ij} - \frac{1}{2}g_{ij}R + \Lambda_{g} g_{ij}.
 \eqn
 ${\cal{L}}_{{M}}$ is the Lagrangian of matter fields, which is a scalar not only with respect to the ${\mbox{Diff}}(M,{\cal{F}})$
 symmetry (\ref{1.4}), but also to  the $U(1)$ symmetry (\ref{2.3}).  
${\cal{L}}_{{V}}$ is an arbitrary Diff($\Sigma$)-invariant local scalar functional
built out of the spatial metric, its Riemann tensor and spatial covariant derivatives, without the use of time derivatives. 
 Assuming that the highest order derivatives are six, and that  the theory  respects 
the parity and time-reflection symmetry   the most general form of  ${\cal{L}}_{{V}}$ is given by  \cite{SVW,KK},  
 \bqn \lb{2.5a} 
{\cal{L}}_{{V}} &=& \zeta^{2}g_{0}  + g_{1} R + \frac{1}{\zeta^{2}}
\left(g_{2}R^{2} +  g_{3}  R_{ij}R^{ij}\right)\nb\\
& & + \frac{1}{\zeta^{4}} \left(g_{4}R^{3} +  g_{5}  R\;
R_{ij}R^{ij}
+   g_{6}  R^{i}_{j} R^{j}_{k} R^{k}_{i} \right)\nb\\
& & + \frac{1}{\zeta^{4}} \left[g_{7}R\Delta R +  g_{8}
\left(\nabla_{i}R_{jk}\right)
\left(\nabla^{i}R^{jk}\right)\right],  ~~~~
 \eqn 
 where the coupling  constants $ g_{s}\, (s=0, 1, 2,\dots 8)$  are all dimensionless, and 
 \bq
 \lb{lambda}
 \Lambda = \frac{1}{2}\zeta^{2}g_{0},
 \eq
 is the cosmological constant. The relativistic limit in the IR, on the other hand, 
 requires, 
 \bq
 \lb{NC}
 g_{1} = -1,\;\; \zeta^2 = \frac{1}{16\pi G} = \frac{M^2_{\text{pl}}}{2},
 \eq
where $G$ is the Newtonian constant.

Note the difference between the notations used here and the ones used in \cite{HMT,Silva} \footnote{In particular, we
have $\varphi = - \nu^{HMT},\; K_{ij} = - K_{ij}^{HMT},\; {\cal{A}} =  a^{HMT},\Lambda_{g} = \Omega^{HMT},\;
{\cal{G}}_{ij} = \Theta_{ij}^{HMT}$, where quantities with super indice ``HMT" are the ones used in \cite{HMT}.}. In this paper, we shall use directly 
the notations and conventions defined in \cite{WM,WW,HW}  without further explanations.    
Then, the field equations are given in Appendix A.

\section{Cosmological  Perturbations }

\renewcommand{\theequation}{3.\arabic{equation}} \setcounter{equation}{0}

In this section, we first give a brief review of the  FRW universe, and then argue that it must be flat in the framework of
the HMT generalization. This is a very important implication. In fact,  one of the main motivations of inflation was to solve
the flatness problem \cite{Inflation}. In the second part of this section,  we  consider scalar
perturbations without restricting ourselves to a particular gauge. We shall closely follow the presentation given in \cite{HW}, which will be referred to as
Paper I. To see the differences, we present our formulas closely parallel to those given in GR \cite{MW09}, and point out the similarities and differences whenever they
raise. 

\subsection{Flatness of the FRW Universe}

 The homogeneous and isotropic universe is described by,
\bq
\lb{3.1}
\bar{N} = 1,\;\; \bar{N}_{i} = 0,\;\;  
\bar{g}_{ij} = a^{2}(t)\gamma_{ij},
\eq
where   
%\bq
%\lb{3.2}
$\gamma_{ij}={\delta_{ij}}{\left(1 + \frac{1}{4}\kappa r^{2}\right)^{-2}}$,
%\eq
with $r^{2} \equiv x^2 + y^2 + z^2,\; \kappa = 0, \pm 1$. As in  Paper I, 
we use symbols with bars to denote the quantities of the background in the ($t, x, y, z$)-coordinates. 
Using the $U(1)$ gauge freedom of Eq.(\ref{2.3}), on the other hand, we can  set
\bq
\lb{3.3}
 \bar{\varphi} = 0.
\eq
Then, we find 
\bqn
\lb{3.4}
\bar K_{ij} &=& - a^{2}H \gamma_{ij}, \;\;\; \bar R_{ij} = 2\kappa\gamma_{ij}, \nb\\
\bar F^{ij}_{A} &=& \frac{2\kappa\bar A}{a^{4}} \gamma^{ij}, \;\;\; \bar F^{ij}_{\varphi} = 0,\;\;\; \bar F^{i}_{\varphi} = 0,\nb\\
\bar F^{ij} &=& \frac{\gamma^{ij}}{a^{2}}\left( - \Lambda + \frac{\kappa}{a^{2}} 
+ \frac{2\Delta_1\kappa^{2}}{a^{4}}  + \frac{12\Delta_2 \kappa^{3}}{a^{6}}\right),
\eqn
and
 \bqn
\lb{3.6}
\bar{\cal{L}}_{K} &=&  3\big(1-3\lambda\big) H^{2},\;\; \bar{\cal{L}}_{\varphi} = 0 = \bar{\cal{L}}_{\lambda}, \nb\\
\bar{\cal{L}}_{A} &=& 2\bar A\Big(\Lambda_{g} - \frac{3 \kappa}{a^{2}}\Big), \nb\\
\bar{\cal{L}}_{V} &=& 2\Lambda - \frac{6 \kappa}{a^{2}} +
\frac{12\Delta_1\kappa^{2}}{a^{4}}  + \frac{24\Delta_2 \kappa^{3}}{a^{6}},
 \eqn
 where $H = \dot{a}/a$ and  
\bq
\lb{3.5}
\Delta_1 \equiv \frac{3g_{2} + g_{3}} {\zeta^{2}},\;\;\; 
\Delta_2 \equiv \frac{9g_{4} + 3g_{5} + g_{6}}{\zeta^{4}}.
\eq
The super-momentum constraint (\ref{eq2}) is satisfied identically, provided that 
$\bar{J}^i = 0$, while the Hamiltonian constraint 
(\ref{eq1}) yields \footnote{Since now the Hamiltonian constraint is a global one, one can include a ``dark matter component as
an integration constant," as first noted in \cite{WOinf}.  
For the sake of simplicity, in this paper we shall not consider this possibility, and it is not difficult to show that our mainly conclusions are
equally applicable to this case.},
 \bq \lb{3.7a}
\frac{1}{2}\big(3\lambda - 1\big)H^{2} + \frac{\kappa}{a^{2}} =
\frac{8\pi G}{3} \bar\rho+ \frac{\Lambda}{3}  
+ \frac{2\Delta_1\kappa^{2}}{a^{4}} + \frac{4\Delta_2 \kappa^{3}}{a^{6}},
 \eq
where $\bar{J}^t \equiv -2\bar\rho$.
 On the other hand, Eqs.(\ref{eq4a}) and (\ref{eq4b}) yield, respectively,
 \bqn
 \lb{3.8a}
 & & H\left(\Lambda_{g} - \frac{\kappa}{a^{2}}\right) = - \frac{8\pi G}{3} \bar J_{\varphi},\\
 \lb{3.8b}
 & & \frac{3\kappa}{a^{2}} -  \Lambda_{g}=  4\pi G \bar J_{A},
 \eqn
while the dynamical equation (\ref{eq3}) reduces to  
 \bqn
 \lb{3.7b}
%& &
 \frac{1}{2}\big(3\lambda - 1\big)\frac{\ddot{a}}{a} &=&  - {4\pi G\over
3}(\bar\rho+3 \bar p)+ {1\over3} \Lambda - \frac{2\Delta_{1}\kappa^{2}}{a^{4}}\nb\\
& &
  - \frac{8\Delta_{2}\kappa^{3}}{a^{6}}
+ \frac{1}{2}\bar A\left(\frac{\kappa}{a^{2}} - \Lambda_{g}\right),
 \eqn
where   $\bar\tau_{ij} =  \bar p\,
\bar g_{ij}$.

The conservation law of the momentum (\ref{eq5b}) is satisfied identically, while 
the one of the energy (\ref{eq5a}) reduces to, 
 \bq \lb{3.8}
\dot{{\bar\rho}} + 3H \left(\bar\rho + \bar p \right) = \bar A \bar J_{\varphi}.
 \eq
 
 From Eqs.(\ref{3.8a}) and (\ref{3.8b}), one can see that when 
 \bq
 \lb{3.9}
 \bar{J}_{A} = 0 =  \bar{J}_{\varphi},
 \eq
 the universe is necessarily flat, $k = 0 = \Lambda_{g}$. This is true for the case where the source is a scalar field \cite{HW}, as
can be seen from Eqs.(\ref{5.8c}) and (\ref{5.8d}) given in the next section, where both  $\bar{J}_{A} $ and $\bar{J}_{\varphi}$ are proportional to 
the spatial gradients of the scalar field $\chi$. This can be easily generalized to the case with multi-scalar fields. 

In general,
the coupling of the gauge field $A$ and the Newtonian prepotential $\varphi$ to a matter field $\psi_{n}$ is given by
\cite{Silva},
\bq
\lb{coupling}
\int{dt d^{3}x \sqrt{g} Z(\psi_{n},  g_{ij}, \nabla_{k})(A - {\cal{A}})},
\eq
where ${\cal{A}}$ is defined as
\bq
\lb{couplinga}
 {\cal{A}} \equiv  - \dot{\varphi} + N^{i}\nabla_{i}\varphi + \frac{1}{2}N \big(\nabla\varphi\big)^{2},
\eq
 and $Z$ is the most general scalar operator under the full symmetry of Eq.(\ref{symmetry}),
with its dimension 
\bq
\lb{DM}
[Z] = 2. 
\eq
For a vector field ($A_{0}, A_{i}$), we have $[A_{0}] = 2,\; [A_{i}] = 0$ \cite{KK}. Then, we find 
\bq
\lb{3.10}
Z(A_{0}, A_{i}, g_{ij}, \nabla_{k}) = {\cal{K}}B_{i}B^{i},
\eq
where ${\cal{K}}$ is an arbitrary function of $A^{i}A_{i}$, and 
\bq
\lb{3.11}
  B_i= \frac{1}{2}\frac{\varepsilon_i^{\;\;jk}}{\sqrt{g}}{\cal{F}}_{jk},\;\;\;
  \nabla^iB_i=0,
  \eq
  with ${\cal{F}}_{ij} \equiv \partial_{j}A_{i} - \partial_{i}A_{j}$.  This can be easily generalized to several vector fields,
  $(A^{(n)}_{0}, A^{(n)}_{i}$), for which we have 
  \bq
\lb{3.12}
Z(\vec{A}_{0}, \vec{A}_{i}, g_{ij}, \nabla_{k}) = \sum_{m,n }{\cal{K}}_{mn}B^{(m)}_{i}B^{(n) i},
\eq
where ${\cal{K}}_{m, n}$ is an arbitrary function of $A^{(k) i}A^{(l)}_{i}$. Then, it is easy to show that in the FRW background, we
have $\bar{J}_{A} = 0$, because $\bar{B}^{(m)}_{i} = 0$ \cite{GMV}, as can be seen from Eq.(\ref{3.11}). With the gauge choice
(\ref{3.3}), one can also show that $\bar{J}_{\varphi} = 0$. Therefore,  an early universe dominated by  vector
fields is also necessarily  flat. This can be further generalized to the case of Yang-Mills fields \cite{CH}. For fermions, on the other hand,
their dimensions are $[\psi_{n}] = 3/2$ \cite{Alex}. Then, $Z(\psi_{n}, g_{ij}, \nabla_{k})$ cannot be a functional of $\psi_{n}$. Therefore, 
in this case $\bar{J}_{A}$ and $\bar{J}_{\varphi}$ vanish identically. 

Although we cannot exhaust all the matter fields, with the special form of the coupling given by Eq.(\ref{coupling}),
it is quite reasonable to argue that {\em the universe is necessarily  flat for all cosmologically viable models in the HMT setup}.
Therefore, in the rest of this paper, we shall consider only the flat FRW universe, i.e., 
\bq
\lb{3.13}
\kappa = 0 = \Lambda_{g},
\eq
for which Eq.(\ref{3.9}) holds.

   \subsection{Linear  Perturbations}
   
As mentioned previously, to solve the strong coupling problem, one needs to impose the condition (\ref{SC}). Once it is satisfied, one can safely carry out
the linear perturbations. With this in mind, as usual,   we study these perturbations in terms of  the conformal time $\eta$, where $\eta = \int{dt/a(t)}$.
Under this coordinate transformation, the fields transform as, 
\bqn
\lb{4.00}
 N &=& a\tilde{N}, \;\;\; N^{i} = a \tilde{N}^{i},\; \;\; g_{ij} = \tilde{g}_{ij},\nb\\
 A &=& a \tilde{A},  \;\;\; \varphi = \tilde{\varphi}, 
 \eqn
 where the quantities with tildes are the ones defined in the coordinates ($t, x^{i}$). With these in mind, 
we write the linear scalar perturbations of the metric in the form,
\bqn
\lb{4.0a}
\delta{N} &=& a \phi,\;\;\; \delta{N}_{i} = a^{2}B_{,i},\nb\\
\delta{g}_{ij} &=& -2a^{2}\big(\psi \delta_{ij} - E_{,ij}\big),\nb\\
A &=& \hat{A} + \delta{A},\;\;\; \varphi = \hat{\varphi} + \delta\varphi,
\eqn
where $\hat{A} = a \bar{A}$ and $\hat{\varphi} = \bar{\varphi}$.  Quantities with hats denote the ones of the background in the
coordinates ($\eta, x^{i}$). 
Under the gauge transformations (\ref{1.4}), they transform as
\bqn
\lb{4.0b}
\tilde{\phi} &=& \phi - {\cal{H}}\xi^{0} - \xi^{0'},\;\;\;
\tilde{\psi} = \psi +  {\cal{H}}\xi^{0},\nb\\
\tilde{B} &=& B +  \xi^{0} - \xi',\;\;\;
\tilde{E} = E -   \xi,\nb\\
\widetilde{\delta\varphi} &=& \delta\varphi - \xi^0 \hat{\varphi}',\;\;\;
\widetilde{\delta{A}} = \delta{A} - \xi^0 \hat{A}' - \xi^{0'} \hat{A}, ~~~
\eqn
where $f = - \xi^0,\; \zeta^i = - \xi^{,i}, \;  {\cal{H}} \equiv a'/a$, and a prime denotes the ordinary derivative with respect 
to $\eta$. Under the $U(1)$ gauge transformations, on the other hand,
we find that
\bqn
\lb{4.0c}
\tilde{\phi} &=& \phi,\;\;\; \tilde{E} = E,\;\;\; 
\tilde{\psi} = \psi,\;\;\;
\tilde{B} =  B - \frac{\epsilon}{a}, \nb\\
\widetilde{\delta\varphi} &=& \delta\varphi + \epsilon,\;\;\;
\widetilde{\delta{A}} = \delta{A} - \epsilon',
\eqn
where $\epsilon = - \alpha$. Then, the  gauge transformations of the whole group $ U(1) \ltimes {\mbox{Diff}}(M, \; {\cal{F}})$
will be the linear combination of the above two. Out of the six unknowns, one can construct three
gauge-invariant quantities \cite{HW},  
\bqn
\lb{4.0d}
\Phi &=& \phi - \frac{1}{a - \hat{\varphi}'}\big(a\sigma - \delta\varphi\big)' \nb\\
& & -  \frac{1}{\big(a - \hat{\varphi}'\big)^2}\big(\hat{\varphi}'' - {\cal{H}}\hat{\varphi}'\big)
\big(a\sigma- \delta\varphi\big),\nb\\
\Psi &=& \psi +  \frac{{\cal{H}}}{a - \hat{\varphi}'}\big(a\sigma- \delta\varphi\big),\nb\\
\Gamma &=& \delta{A} +\Bigg[\frac{a\big(\delta\varphi - \hat{\varphi}'\sigma\big) - \hat{A}\big(a\sigma - \delta\varphi\big)}{a - \hat{\varphi}'}\Bigg]',
\eqn
where  $\sigma\equiv E'-B$.  
For the background, we have chosen the gauge (\ref{3.3}), for
which Eq.(\ref{4.0d}) reduces to
\bqn
\lb{4.0db}
\Phi &=& \phi  - \frac{1}{a}\big(a\sigma - \delta\varphi\big)' , \nb\\ 
\Psi &=& \psi - \frac{{\cal{H}}}{a}\big(\delta\varphi -a\sigma\big),\nb\\
\Gamma &=& \delta{A} + \Bigg[\delta\varphi - \frac{\hat{A}}{a}\big(a\sigma -  \delta\varphi\big)\Bigg]',\; (\bar{\varphi} = 0).
\eqn

Then, for  the general perturbations (\ref{4.0a}), we have 
 \bqn
 \lb{4.0dc}
 \delta{K}_{ij} &=& a\Big\{\psi' \delta_{ij} - \sigma_{,ij}\nb\\
 && ~~~~~ + {\cal{H}}\Big[\left(2\psi + \phi\right)\delta_{ij} - 2 E_{,ij}\Big]\Big\}, \nb\\
 \delta{R}_{ij} &=& \psi_{,ij} + \partial^2\psi \delta_{ij}. 
 \eqn
Thus, to first-order  the Hamiltonian and momentum constraints become,  respectively,
 \bqn \lb{4.4}
&&  \int d^{3}x\Bigg\{\partial^2\psi
- \frac{1}{2}\big(3\lambda -1\big){\cal H}
\Big[3(\psi'+{\cal H} \phi)-\partial^2\sigma\Big] \nb\\ 
&&   ~~~~~~~~~~~~ -{4\pi G a^{2}}\delta{\mu}\Bigg\}=0,\\
 \lb{4.5}
&&  (3\lambda -1)\big({\psi}'  + {\cal{H}}\phi\big)     
 + (1- \lambda)\partial^2\Big(\sigma  - \frac{1}{a}\delta\varphi\Big) \nb\\
 && ~~~~~~~~~~~~~~~~~~~~~~~~ = 8\pi G a {q} + \Delta(\eta), 
 \eqn
where 
\bq
\lb{4.4a}
\delta\mu \equiv -\frac{1}{2}\delta{J^{t}}, \;\;\;
 \delta{J}^{i} \equiv \frac{1}{a^{2}} q^{,i}, 
 \eq
 $q^{,i} = \delta^{ij}q_{,j}$, and $\Delta(\eta)$ is an integration function. In GR, it is usually set to zero   
 \cite{MW09}. However, in the present case,  since    
$\phi = \phi(\eta)$,  another interesting choice is  
$\Delta(\eta) =   (3\lambda -1) {\cal{H}}\phi$,  which  will cancel  the second term
in the left-hand side of Eq.(\ref{4.5}). 

On the other hand, the linearized equations (\ref{eq4a}) and (\ref{eq4b}) reduce, respectively,  to 
 \bqn
 \lb{4.6a}
& & 2{\cal{H}}\partial^2\psi +  \big(1-\lambda\big)\partial^{2}\Bigg[3\big(\psi' + {\cal{H}}\phi\big) %& & ~~~~~~~~~~  
- \frac{1}{a} \partial^2\Big(a\sigma  - \delta\varphi\Big)\Bigg]   \nb\\
&& ~~~~~~~~~~~~~~~~~~~~~~~~~ = 8\pi G a^{3} \delta J_{\varphi},\\
 \lb{4.6b}
&&  \partial^2\psi = 2\pi G a^{2} \delta J_{A},
 \eqn
 while  the linearly perturbed dynamical equations can be divided into the trace and traceless parts. The trace part  reads,
 \bqn
\lb{4.7a}
& &  \psi'' + 2{\cal{H}}\psi'  + {\cal{H}}\phi' + \big(2{\cal{H}}' + {\cal{H}}^{2}\big)\phi \nb\\
&&  ~~~~ - \frac{1}{3}\partial^{2}\big(\sigma' + 2{\cal{H}}\sigma\big)\nb\\ 
& & ~~~~
 - \frac{2}{3(3\lambda-1)}\Bigg(1 + \frac{\alpha_{1}}{a^{2}}\partial^{2}
   + \frac{\alpha_{2}}{a^{4}}\partial^{4}\Bigg)\partial^{2}\psi \nb\\
   & & ~~~~ - \frac{\Lambda_g a}{3(3\lambda-1)}2\hat{A}\left( \partial^2 E - 3\psi - 3 \phi/2\right) \nb\\
   & & ~~~~ - \frac{\Lambda_{g}a}{3(3\lambda - 1)}\Big[2\hat{A}\big(3\psi - \partial^{2}E\big) + 3\big(\delta\varphi' + \delta{A}\big)\Big]\nb\\
   & & ~~~~ + \frac{2}{3(3\lambda - 1)a}\partial^{2}\big(\hat{A}\psi - \delta{A}  +  {\cal{H}}\delta\varphi\big)\nb\\ 
   & & ~~~~ + \frac{\lambda - 1}{(3\lambda - 1)a}\partial^{2}\big(\delta\varphi' +  {\cal{H}}\delta\varphi\big) 
                         = \frac{8\pi G a^{2}}{3\lambda-1}\delta{p}, ~~~~
\eqn 
where  
 \bqn
 \lb{4.8} 
 \alpha_{1} &\equiv& \frac{8g_{2} + 3g_{3}}{\zeta^{2}},\;\;\;
   \alpha_{2} \equiv  \frac{8g_{7}-3g_{8}}{\zeta^{4}},\nb\\  
\delta{p} &\equiv& \delta{\cal{P}} + \frac{2}{3}\bar{p}\partial^{2} E,\;\;\;
\Pi^{GR} \equiv \Pi + 2\bar{p}E,\nb\\
  \delta{\tau}^{ij} &=& \frac{1}{a^{2}}\Big[\big(\delta{\cal{P}} + 2\bar{p}\psi\big)\delta^{ij} + \Pi^{,<ij>}\Big],\nb\\
\Pi^{,<ij>} &=& \Pi^{, ij} - \frac{1}{3}\delta^{ij} \partial^{2}\Pi.
 \eqn 
 The traceless part   is given by
 \bqn
 \label{4.7b}
&&  \psi - \phi + \sigma' + 2{\cal H}\sigma  %- \phi   
+ \frac{1}{a^{2}}\Big(\alpha_{1} + \frac{\alpha_{2}}{a^{2}}\partial^{2}\Big)\partial^2\psi
\nb\\
& & ~~~ %- 2 \hat{A} \Lambda_g a^ 2 E 
- \frac{1}{a}\Big[\hat{A}\psi  - \big(\delta{A} - {\cal{H}}\delta\varphi\big)\Big] %\nb\\ & & ~~~~~~~~~~%~~~~~~~~~~~~~~~~~
=8\pi G a^{2} \Pi^{GR} + G(\eta),\nb\\
 \eqn
where $G(\eta)$ is another integration function. Again,  in GR it is set to zero \cite{MW09}.
%, where  $\lambda = 1, \; g_{i > 1} = 0$, and $A = \varphi = 0$, one needs to choose $G(\eta) = 0$. 
But, similar to the momentum constraint (\ref{4.5}), one can also choose $G(\eta) = -\phi$ 
so that   the second term in the left-hand side of the above equation is canceled.

 The conservation laws (\ref{eq5a}) and (\ref{eq5b}) to first order are  given, respectively,  by,  
 \bqn \lb{4.9a}
& & \int  d^{3}x \Bigg\{2a\Big[\delta\mu' + 3{\cal H}
\left(\delta{\cal P} + \delta\mu\right) +2\bar{p} {\cal H}\partial^2 E  \nb\\
& & ~~~~~~~~~~ + \left(\bar\rho + \bar p\right)\left(\partial^2 E-3{\psi}\right)'-\bar{J}_\varphi \delta \varphi'\Big]  \nb\\
& & ~~~~~~~~~~  - \hat{A}\big({\delta J_A}'+3{\cal H}  {\delta J_A}\big)  - 3{\cal{H}}\bar{J}_{A}\big(\delta{A} - \hat{A}\phi\big)\nb\\
& & ~~~~~~~~~~ + \hat{A}\bar{J}_{A}\big(3\psi - \partial^2 E\big)'
\Bigg\} =0,\\
\lb{4.9b}
 & & q'+3 {\cal H}q  - a\delta{p}   - {2a\over3} \partial^2\Pi^{GR} %\nb\\
%&& ~~~~~~~~~~
%+ \frac{1}{2}\bar{J}_{A} \left(\delta{A} - 3{\cal H} \delta\varphi\right) 
= I(\eta), ~~
 \eqn
where %$\bar{J}_{A}$  is given by Eq.(\ref{3.8b}), and 
$I(\eta)$ is another integration function of $\eta$ only. In GR, it is usually chosen to be zero
%$I(\eta) = 0$
 \cite{MW09}.

This completes the general description of linear scalar perturbations in the flat FRW background 
 in the framework of the HMT setup with any given $\lambda$ \cite{Silva}, without choosing  any specific gauge for the linear perturbations.
 However, before closing this section, let us consider some possible gauges. 
 
 \subsection{Gauge Choices}

To consider the gauge choices, we first note that  
$$
\xi = \xi(\eta, x), \; \xi^{0} = \xi^{0}(\eta).
$$
 Then, from Eqs.(\ref{4.0b}) and (\ref{4.0c}) one immediately finds that {\em the spatially flat gauge} $\tilde\psi = 0 = \tilde{E}$
\cite{MW09}
is impossible in the HMT setup. Since $\phi = \phi(\eta)$, a natural gauge  for the time sector  is 
\bq
\lb{phi}
\tilde{\phi} = 0,
\eq
for which $\xi^{0}$ is uniquely fixed up to a constant $C$, 
\bq
\lb{6.16a}
\xi^{0}(\eta) = \frac{1}{a(\eta)}\int^{\eta}{a(\eta')\phi(\eta')d\eta'} + \frac{C}{a(\eta)}.
\eq
%where $C$ is an integration constant. 
Then, depending on the choices of $\xi$ and $\epsilon$, we can have various different gauges.

\subsubsection{Longitudinal Gauge}

The longitudinal gauge in GR is defined as \cite{MW09},
\bq
\lb{6.16b}
\tilde{E} = 0 = \tilde{B},
\eq
which is impossible in the HL theory without the U(1) symmetry \cite{WM}. However, with the U(1) gauge freedom,  Eqs.(\ref{4.0b}) and (\ref{4.0c})
show that now this gauge becomes   possible with the choice,  
\bq
\lb{6.16c}
\xi = E, \;\;\; \epsilon = a(B  - E' + \xi^{0}),
\eq
where $\xi^{0}$ is given by Eq.(\ref{6.16a}). It should be noted that this gauge is fundamentally different from that
given in  GR   \cite{MW09}, because now we also have $\tilde{\phi} = 0$.

 \subsubsection{Synchronous Gauge}
 
 In GR,  the synchronous gauge is defined as \cite{MW09},
 \bq
 \lb{6.16d}
 \tilde{\phi} = 0 = \tilde{B}.
 \eq
 However, this is already implied  in   the above longitudinal gauge. With the extra $U(1)$ gauge freedom $\epsilon$, we can further require,
 \bq
 \lb{6.16da}
 (i) \;  \widetilde{\delta\varphi} = 0,\;\;\;
 {\mbox{or}} \;\;\; (ii) \;  \widetilde{\delta{A}} = 0. 
 \eq
 The former will be referred to as {\em the Newtonian synchronous gauge}, while the latter {\em the Maxwell synchronous gauge}.
 For the  Newtonian synchronous gauge, $\epsilon$ and $\xi$ are given by
 \bqn
 \lb{6.16e}
 \xi(\eta, x) &=& \int^{\eta}{\left[B + \xi^{0} + \frac{1}{a}\left(\delta\varphi -  \xi^{0}\hat{\varphi}'\right)\right]d\eta'} + D(x),\nb\\
  \epsilon(\eta, x) &=& \xi^{0}\hat{\varphi}' - \delta\varphi,
  \eqn
  where $D(x)$ is an arbitrary function of $x^{i}$ only.  For the Maxwell synchronous gauge, they are given by
   \bqn
 \lb{6.16f}
  \epsilon(\eta, x) &=&  \int^{\eta}{\left[\delta{A} - \left(\xi^{0}\hat{A}\right)'\right] d\eta'} + D_{1}(x),\nb\\
 \xi(\eta, x) &=& \int^{\eta}{\left(B + \xi^{0} - \frac{\epsilon}{a}\right)d\eta'} + D_{2}(x),
  \eqn
  where $D_{1}(x)$ and $D_{2}(x)$ are other two arbitrary functions of $x^{i}$ only. From the above one can see that none of them can fix the gauge uniquely.

\subsubsection{Quasilongitudinal Gauge}

 In \cite{WW,HW}, the   gauge,   
  \bq
  \lb{6.16}
\tilde{\phi} = \tilde{E} =  \widetilde{\delta\varphi} = 0,
 \eq
 was used.  With this gauge, we find that  
 \bq
 \xi(\eta, x) = E(\eta, x),\;\;\; \epsilon(\eta, x) = \xi^{0}\hat{\varphi}' - \delta\varphi(\eta, x),
 \eq
 and $\xi^{0}$ is given by  Eq.(\ref{6.16a}). Thus, in this case the gauge freedom of Eqs.(\ref{4.0b}) and (\ref{4.0c}) are also uniquely determined up to the constant $C$, 
 similar to the longitudinal gauge (\ref{6.16b}). 
 
 Note that instead of choosing the above gauge, one can also choose  
 \bq
 \lb{6.16g}
\tilde{\phi} = \tilde{E} =  \widetilde{\delta{A}} = 0,
 \eq
for which we have 
 \bqn
 \xi(\eta, x) &=& E(\eta, x),\nb\\ 
   \epsilon(\eta, x) &=&  \int^{\eta}{\left[\delta{A} - \left(\xi^{0}\hat{A}\right)'\right] d\eta'} + D_{3}(x), 
 \eqn
 where $D_{3}(x)$ is another integration function of $x^{i}$ only. Thus, unlike the gauge (\ref{6.16}),  now the gauge is fixed only up to a constant $C$ and an arbitrary function $D_{3}(x)$. 
 
To be distinguish from the one defined in the case without the U(1) symmetry \cite{WM}, we shall refer the gauge (\ref{6.16}) to as {\em the Newtonian quasilongitudinal gauge}, and
Eq.(\ref{6.16g})  {\em the Maxwell quasilongitudinal gauge}.

\section{Inflation of a scalar field}

\renewcommand{\theequation}{4.\arabic{equation}} \setcounter{equation}{0}

% in the HMT setup. 
In Appendix B, we   construct the action for  a single scalar field. In this section, we apply the perturbations developed in Sec. III to study inflationary models of such a scalar field.  
To this goal, let us first consider the slow-roll conditions.

\subsection{Slow-Roll Inflation}
 
 For the flat FRW background,  we find that
 \bqn
 \lb{5.12}
 \bar{J}^{t} &=& -2f\left(\frac{1}{2} \dot{\bar\chi}^{2}  + \tilde{V}(\bar\chi)\right) \equiv -2\bar{\rho},\nb\\
  \bar{J}^{i} &=& \bar{J}_{\varphi} = \bar{J}_{A} =  0,\nb\\
  \bar{\tau}_{ij} &=& f a^2  \left(\frac{1}{2} \dot{\bar\chi}^2 -\tilde{V}(\bar\chi)\right) \delta_{ij} \equiv a^2 \bar{p} \delta_{ij},
   \eqn 
where $\tilde{V}(\bar\chi) \equiv  V(\bar\chi)/f$. Then,  Eqs.(\ref{3.7a}) - (\ref{3.8b}) and (\ref{3.8}) yield $\Lambda_{g} = 0$ and 
 \bq
  \lb{5.13a}
  H^{2}   = \frac{8\pi \tilde{G}}{3} \left(\frac{1}{2}\dot{\bar\chi}^{2}+ \tilde{V} (\bar\chi)\right)  + \frac{\tilde{\Lambda}}{3}, ~~~\\
  \eq
 where
 \bq
 \lb{5.13c}
 \tilde{G} \equiv \frac{2fG}{3\lambda -1},\;\;\; \tilde{\Lambda} \equiv \frac{2\Lambda}{3\lambda - 1}.
 \eq
On the other hand,  Eq.(\ref{5.10}) reduces to,
 \bq
 \lb{5.14}
 \ddot{\bar\chi} + 3 H \dot{\bar\chi} + {\tilde{V}'}= 0.
 \eq
 Eqs.(\ref{5.13a}) and (\ref{5.14}) are identical to these given in GR \cite{MW09}, if one identifies $\tilde{G}$ and $\tilde{\Lambda}$ 
 to the Newtonian  and cosmological constants, respectively. As a result, all the conditions for inflationary models  obtained in GR
 are equally applicable to the current case, as long as the background is concerned. In particular, the slow-roll conditions,
 \bq
 \lb{5.14a}
\tilde \epsilon_{{\scriptscriptstyle V}} ,\; |\tilde\eta_{{\scriptscriptstyle V}}| \ll 1,
 \eq
need to be imposed    in order to get enough e-fold,  where
 \bqn
 \lb{5.14b}
\tilde \epsilon_{{\scriptscriptstyle V}} &\equiv& \frac{\tilde{M}_{\text{pl}}^{2}}{2}\left(\frac{{\tilde V}'}{\tilde V}\right)^{2} = \frac{3 \lambda -1}{2 f}\epsilon_{{\scriptscriptstyle V}} ,\nb\\
\tilde \eta_{{\scriptscriptstyle V}}  &\equiv& \tilde{M}_{\text{pl}}^{2} \left(\frac{{\tilde V}''}{ \tilde V}\right)= \frac{3 \lambda -1}{2 f}\eta_{{\scriptscriptstyle V}},
 \eqn
 with $\tilde{M}_{\text{pl}}^{2} \equiv 1/(8\pi \tilde{G})$, and $\epsilon_{{\scriptscriptstyle V}}$ and $\eta_{{\scriptscriptstyle V}}$ are the ones defined in GR \cite{Inflation}.
 
 However, due to the presence of high-order spatial derivatives, the perturbations will be dramatically different, as to be shown below.

 \subsection{Linear Perturbations}
 
In this section,  in order for the formulas developed below to be applicable to as many cases as possible, we shall not restrict ourselves to any specific gauge. %,.
Then, to first-order  we  find that
 \bqn
 \lb{5.15a}
 \delta\rho &\equiv&  \delta\mu  =   \frac{f \bar{\chi}'}{a^{2}} \big(\delta \chi' - \bar{\chi}' \phi\big)  +  \frac{V_4}{a^{4}} \partial^4\delta \chi + {V'} \delta \chi, \nb\\
  q &=& \frac{f \bar{\chi}'}{a}\delta \chi,\;\;\;  
  \delta J_A = \frac{2c_{1}}{ a^{2}} \partial^2 \delta \chi, \nb\\
  \delta J_{\varphi} &=& \frac{1}{a^{3}} \Big[\big({c}'_{1} \bar{\chi}'+ {c}_{1}{\cal H} - f \bar{\chi}'\big)\partial^2 \delta \chi + {c}_1 \partial^2 \delta \chi' \Big],\nb\\
   \delta p &=&   \frac{f \bar{\chi}' }{a^2}(\delta \chi' -\bar{\chi}' \phi) -{V}' \delta \chi, \nb\\
\Pi &=&-2 \bar{p} E ,  ~~~~~ \Pi^{GR} = 0. 
 \eqn
 Hence, Eqs. (\ref{4.4}) - (\ref{4.6b})  read, respectively,
 \bqn
 \lb{5.17a}
 &&  \int d^{3}x\Bigg\{\partial^2\psi
- \frac{1}{2}\big(3\lambda -1\big){\cal H}
\Big[3(\psi'+{\cal H} \phi)-\partial^2\sigma\Big]\Bigg\} \nb\\ 
&&   ~~~~ =\int d^{3}x4\pi G\Bigg\{f\bar{\chi}' \Big(\delta{\chi}' - \bar{\chi}' \phi\Big) + \frac{V_{4}}{a^{2}} \partial^4 \delta \chi \nb\\
&& ~~~~~~~~~~~~~~~~~~~~~~~~~ + a^2{V}' \delta \chi\Bigg\},\\
 \lb{5.17aa}
&& (3\lambda -1){\psi}'    + (1- \lambda)\partial^2\Big(\sigma  - \frac{1}{a}\delta\varphi\Big) \nb\\
&&~~~~~~~~~~~~~~~~~~~~~~~~~ = 8\pi G f \bar{\chi}' \delta \chi, \\
 \lb{5.17ab}
& & 2{\cal{H}}\psi +  \big(1-\lambda\big)\Bigg[3\psi'   - \frac{1}{a} \partial^2\Big(a\sigma  - \delta\varphi\Big)\Bigg] 
\nb\\
&&   ~~~~~ = 8\pi G\Big[\big({c}'_{1} \bar{\chi}'+ {c}_{1}{\cal H} - f \bar{\chi}'\big) \delta \chi + {c}_1  \delta \chi'\Big] ,\\
 \lb{5.17ac}
&& \psi = 4\pi G {c}_1  \delta \chi.
 \eqn
Note that in writing Eq.(\ref{5.17aa}), we had chosen $\Delta(\eta) =   (3\lambda -1) {\cal{H}}\phi$.
 It is also interesting to note that, unlike the case without the U(1) symmetry \cite{WWM}, now the metric perturbation $\psi$ is proportional to $\delta\chi$. 
 It is this difference that leads to a master equation, as to be shown below. Without the U(1) symmetry, this is in general impossible \cite{WWM}. 
 
 The trace and traceless parts of dynamical equation read, respectively, 
 \bqn
 \lb{5.17b}
 & &  \psi'' + 2{\cal{H}}\psi'  + {\cal{H}}\phi' + \big(2{\cal{H}}' + {\cal{H}}^{2}\big)\phi - \frac{1}{3}\partial^{2}\big(\sigma' + 2{\cal{H}}\sigma\big)\nb\\ 
& & ~~~~
 - \frac{2}{3(3\lambda-1)}\Bigg(1 + \frac{\alpha_{1}}{a^{2}}\partial^{2}
   + \frac{\alpha_{2}}{a^{4}}\partial^{4}\Bigg)\partial^{2}\psi \nb\\
   & & ~~~~ + \frac{2}{3(3\lambda - 1)a}\partial^{2}\big(\hat{A}\psi - \delta{A}  +  {\cal{H}}\delta\varphi\big)\nb\\ 
   & & ~~~~ + \frac{\lambda - 1}{(3\lambda - 1)a}\partial^{2}\big(\delta\varphi' +  {\cal{H}}\delta\varphi\big)  \nb\\
   & & ~~~~  = \frac{8\pi G }{3\lambda-1}\Bigg[ f\bar{\chi}' (\delta \chi' -\bar{\chi}' \phi) - a^2{V}'  \delta \chi \Bigg], \\
  \lb{5.17c}
 &&  \psi  + \sigma' + 2{\cal H}\sigma  %- \phi   
+ \frac{1}{a^{2}}\Big(\alpha_{1} + \frac{\alpha_{2}}{a^{2}}\partial^{2}\Big)\partial^2\psi\nb\\
& & ~~~ -\frac{1}{a}\Big[\hat{A}\psi  - \big(\delta{A} - {\cal{H}}\delta\varphi\big)\Big] =0,
 \eqn
where in writing Eq.(\ref{5.17c}) we had set $G(\eta) = - \phi$.  The energy conservation law now takes the form,
 \bqn
 \lb{5.18}
 && \int d^3 x a^2 \bar{\chi}' \Bigg\{f \delta \chi'' + 2 {\cal H} f \delta \chi' + a^2 {V}'' \delta \chi + 2 a^2 \bar{V}'\phi\nb\\
 && ~~~ -f \bar{\chi}' \Big[\phi - \left(\partial^2 E - 3\psi\right)\Big]'-\frac{\bar{A}}{\bar{\chi}'}\frac{\left(a c_1 \delta \chi\right)'}{a}\Bigg\}\nb\\
 &&= -\int d^3 x \; \partial^4\Bigg\{{V}_{4}  \delta \chi' + \Big({V}'_{4} \bar{\chi}' - {V}_{4} {\cal H}\Big)  \delta \chi\Bigg\},
 \eqn
The momentum conservation is identically satisfied,
while the Klein-Gordon equation becomes
 \bqn
 \lb{5.19}
 && f \Bigg\{\delta \chi'' + 2 {\cal H} \delta \chi' - \bar{\chi}' \left[3 \psi' + \phi' - \partial^2 \sigma\right]\Bigg\} \nb\\
 && ~~~~~~ + 2a^2 V' \phi + \left(a^2V''-\partial^2\right)\delta\chi\nb\\
 && ~~ = \frac{\partial^2}{a} \big[2\hat{A}\left({c}'_{1} - {c}_{2}\right)  \delta \chi - c_{1} \delta\varphi' + f \bar{\chi}' \delta \varphi + c_1 \delta{A}\big]\nb\\
 && ~~~~~~ + 2 \Big({V}_{1} -\frac{{V}_{2}+{V}'_{4}}{a^2}\partial^2 -\frac{{V}_{6}}{a^4} \partial^4\Big)\partial^2 \delta \chi,
 \eqn
which can be   rewritten as a perturbed energy balance equation,
\bqn
\lb{5.20}
&& \delta \rho' +  3 {\cal{H}} (\delta \rho + \delta p) \nb\\
&& ~~~~~~  - (\bar{\rho}+ \bar{p})\left(3\psi'-\partial^2\sigma- \frac{1}{f}\partial^2 (v+B)\right)\nb\\
&& = \frac{1}{f} (\bar{\rho}+ \bar{p}) \delta Q^{\text{HMT}},
\eqn
where
\bqn
\lb{5.21}
\delta Q^{\text{HMT}}  &=& \frac{{V}_{4}}{a^2 \bar{\chi}'^2} \partial^4 \delta \chi'    + \frac{1}{\chi'}\Bigg[2 {V}_{1} -2 \frac{{V}_{6}}{a^4} \partial^4 \nb\\
& & -\frac{1}{a^2}\left(2{V}_{2}+{V}'_{4}+ \frac{{V}_{4}{\cal{H}}}{\bar{\chi}'}\right)\partial^2 \Bigg]\partial^2 \delta \chi \nb\\
 &&  +\frac{1}{a\bar{\chi}'} \partial^2\Big[2\hat{A}\left({c}'_{1} - {c}_{2}\right)  \delta \chi - c_{1} \delta\varphi' \nb\\
 & & + f \bar{\chi}' \delta \varphi + c_1 \delta{A} \Big],\nb\\
  q &\equiv& -a(\bar{\rho}+\bar{p})(v+B).
\eqn

\subsection{Uniform Density Perturbation}

Under the gauge transformations (\ref{4.0b}) and (\ref{4.0c}), $\delta\chi$ and $\delta\rho$ transforma, respectively, as
\bq
\lb{6.1a}
\widetilde{\delta\chi} = \delta\chi - \xi^{0}\bar\chi',\;\;\;
\widetilde{\delta\rho} = \delta\rho - \xi^{0}\bar\rho'.
\eq
Therefore, the  quantity $\zeta$ defined by
\bq
\lb{6.2}
- \zeta \equiv \psi + \frac{{\cal{H}}}{\bar{\rho}'} \delta \rho,
\eq
is  gauge-invariant. In GR it is often referred to as the  gauge-invariant perturbation on uniform-density hypersurfaces. 
 It can be shown that it obeys the evolution equation,
\bqn
\lb{6.3}
\zeta' = -\frac{{\cal{H}} \delta p_{nad}}{\bar{\rho}+\bar{p}}  + \frac{1}{3} \left[\delta Q^{\text{HMT}} - \partial^2 \sigma -\frac{\partial^2}{f} (v+B)\right],~~~~
\eqn
where the non-adiabatic pressure perturbation is defined as
\bq
\lb{6.4}
\delta p_{nad} \equiv \delta p -\frac{\bar{p}'}{\bar{\rho}'}\delta \rho = \delta p^{\text{GR}}_{nad}+\delta p^{\text{HMT}}_{nad},
\eq
with,
\bqn
\lb{6.5}
 \delta p^{\text{GR}}_{nad} &\equiv&\frac{2}{3a^2}\left(2+ \frac{ \bar{\chi}''}{{\cal{H}}\bar{\chi}'}\right)\Big[\bar{\chi}' \left(\delta \chi' -\bar{\chi}'\phi\right)\nb\\
 &&  -(\bar{\chi}''-{\cal{H}}\bar{\chi}') \delta \chi\Big]\nb\\
&=& - \frac{2 a \tilde{V}'}{3{\cal{H}}\bar{\chi}'}\left[\left(\frac{\bar{\chi}'}{a}\right)\left(\delta\chi' - \bar{\chi}'\phi\right)- \left(\frac{\bar{\chi}'}{a}\right)'\delta\chi\right], \nb\\
 \delta p^{\text{HMT}}_{nad} &\equiv& \left(1+\frac{2 \bar{\chi}''}{{\cal{H}}\bar{\chi}'}\right)\frac{{V}_{4}}{3a^4}\partial^4 \delta \chi .
\eqn
Note that Eq.(\ref{6.3}) is quite similar to that of the case without the U(1) symmetry   \cite{WWM}, and the only   difference is the inclusion of the  $U(1)$ 
gauge field $A$ and Newtonian prepotential $\varphi$ in   $\delta Q^{HMT}$,  
as one can see from Eq.(\ref{5.21}) given above and Eq.(4.3) given in \cite{WWM}. But, these terms vanish in the super-horizon region. As a result, all the conclusions
obtained in \cite{WWM} in this region are equally applicable to the present case. In particular, the perturbations in this region are adiabatic during the slow-roll inflation,
as in GR.

 \subsection{Comoving Curvature Perturbation}
 
 On the other hand, the comoving curvature perturbation, defined by
 \bq
 \lb{6.6}
 {\cal{R}}  = \psi + \frac{\cal{H}}{\bar{\chi}'} \delta\chi,
 \eq
 is  gauge-invariant even in the HL theory. From its definition, it can be shown that  ${\cal{R}}$  satisfies the equation,
 \bq
 \lb{6.7}
  {\cal{R}}'  =  {\cal{H}}  {\cal{S}}  + \frac{ {\cal{H}}'-  {\cal{H}}^2}{\bar{\chi}'}\delta \chi + \psi' +{\cal{H}}\phi,
 \eq
 where the dimensionless intrinsic entropy perturbation $ {\cal{S}}$ is defined as
 \bq
 \lb{6.8}
 {\cal{S}} \equiv \frac{\delta \chi'-\bar{\chi}'\phi}{\bar{\chi}'} - \frac{\bar{\chi}'' - {\cal{H}} \bar{\chi}'}{\bar{\chi}'^2} \delta \chi
  =  - \frac{3{\cal{H}}} {2 \tilde{V}' \bar{\chi}' } \delta p^{\text{GR}}_{nad},
 \eq
where to get the last step  Eq.(\ref{6.5}) was used.  
In terms of  ${\cal{R}}$ the super-momentum constraint (\ref{5.17aa})  can be written in the form, 
 \bq
 \lb{6.10}
  {\cal{R}}'  =  {\cal{H}}  {\cal{S}}  + \frac{\lambda -1}{3 \lambda -1} \partial^2 \left(\sigma - \frac{\delta \varphi}{a}\right),
 \eq
which reduces to   ${\cal{R}}'  =  {\cal{H}}  {\cal{S}}$ on all scales   in the relativistic  limit $\lambda \rightarrow 1$.
Thus,   in the slow-roll approximations  and neglecting the spatial gradients on large scales, we obtain  the same conclusion as that given in \cite{WWM}, namely, 
the comoving curvature perturbation has two modes on large scales,     a constant mode and a rapidly decaying mode,  given by
  \bq
  \lb{6.28}
 {\cal{R}} \simeq C_{1} +C_{2} \int \frac{d\eta}{a^2}.
 \eq
In addition, unlike that in GR where the local Hamiltonian constraint enforces  adiabaticity on large scales, in HMT setup it is the slow-roll evolution ($\ddot{\bar{\chi}} =0$, or, 
$\bar{\chi}''={\cal{H}}\bar{\chi}'$) that  leads to rapidly decaying entropy perturbations at late times.

Note that we could also find the first-order equation for $  {\cal{S}} $ by using the Klein-Gordon equation, which can be written in the
 form, 
 \bqn
 \lb{6.11}
&&  {\cal{S}}' + \left( 2\frac{\bar{\chi}''}{\bar{\chi}'}+{\cal{H}}\right) {\cal{S}} = \frac{1}{f\bar{\chi}'} \Bigg\{\left[f\bar{\chi}' 3 {\cal{H}}\phi\right] \nb\\
&&  +  \partial^2\left[\left(1+2 {V_1}\right)\delta \chi - \frac{ 2 f\bar{\chi}'}{3 \lambda - 1}(\sigma - \frac{\delta \varphi}{a} ) \right]\nb\\
&&  + \frac{\partial^2}{a}\left[2 \hat{A} \left({c_1}'-{c_2}\right) \delta \chi+ c_1 \delta A - {c_1}\delta \varphi' \right] \nb\\
&& - 2 \frac{\partial^4}{a^2} \left[\left({V_2} +{V_4}'\right)+\frac{{V_6}}{a^2}\partial^2\right] \delta \chi  \Bigg\}.
 \eqn
Thus, in the large scales (neglecting all the spatial gradient terms), the first term in the right-hand side is function of $\eta$ only
 (Recall that $\phi = \phi(\eta)$). Then, the corresponding entropy equation depends only on time  on these large-scales.

\section{ Scalar   perturbations in Sub- and Super-Horizon Scales}  

\renewcommand{\theequation}{5.\arabic{equation}} \setcounter{equation}{0}

So far, we have not chosen any gauge. In this section, we shall   restrict ourselves to the 
Newtonian quasilongitudinal gauge defined by Eq.(\ref{6.16}) in Sec. III.C, i.e., 
\bq
\lb{NLG}
\phi =  E = \delta\varphi = 0.
\eq
Then,   Eqs.(\ref{5.17a}) - (\ref{5.19}) can be cast in the forms of Eqs.(\ref{6.17a}) - (\ref{6.17h}). 
 From Eqs.(\ref{6.17c}), (\ref{6.17e}), (\ref{6.17g}), we can express $\psi, \; B$ and $ \delta{A}$ in terms of $\delta\chi$, 
 and then submit them into Eq.(\ref{6.17h}), we obtain a master equation for $\delta\chi$, 
 which can be written as
 \bq
 \lb{6.38}
 \delta \chi'' + {\cal{P}} \delta \chi' + {\cal{Q}} \delta \chi = {\cal{F}} \partial^2\delta \chi,
 \eq
where
\bqn
\lb{6.39}
\beta_0 &\equiv& f+ \frac{4 \pi G {c_1}^2}{|c_{\psi}^{2}|},\nb\\
{\cal{P}} &\equiv& \frac{1}{\beta_0} \left(\beta'_0 + 2 {\cal{H}} \beta_0\right),\nb\\
%\delta_2 &\equiv& 2{\cal{H}}f +\frac{4\pi G {c_1}^2}{|c_{\psi}^{2}|}\Big[2{\cal{H}} +  2\frac{c_1' \bar{\chi}'}{c_1}\Big], \nb\\
{\cal{Q}} &\equiv& \frac{1}{\beta_0}\Bigg\{a^2 V'' + \frac{4 \pi G c_1 c''_1 \bar{\chi}'^2 }{|c_{\psi}^{2}|} -\frac{8 \pi G}{\lambda -1}f\bar{\chi}'^2\Big(f- {c_1}'\Big)\nb\\
&& -4 \pi G c_1 a^2 V' \left(3 +\frac{ c'_1}{f |c_{\psi}^{2}|} - \frac{1}{|c_{\psi}^{2}|}\right)\Bigg\},\nb\\
{\cal{F}}&\equiv& \frac{1}{\beta_0}\Bigg\{1 + 2 V_1 + 2\bar{A}({c_1}'-{c_2}) - 4 \pi G {c_1}^2 \left(1-\bar{A}\right) \nb\\
&& ~~~~~~~ -\frac{2}{a^2} \Big(V_2 +V'_4 + 2 \pi G  \alpha_1{c_1}^2\Big)\partial^2 \nb\\
&& ~~~~~~~ -\frac{2}{a^4} \Big(V_6 + 2 \pi G  \alpha_2 {c_1}^2 \Big)\partial^4\Bigg\},
\eqn
with
\bq
\lb{6.39a}
c_{\psi}^{2} \equiv \frac{\lambda - 1}{1- 3\lambda}.
\eq
Setting 
\bq
\lb{6.40}
\delta\chi = \exp \left(-\frac{1}{2} \int {\cal{P}} d\eta \right) u,
\eq
 one can write Eq.(\ref{6.38}) in the momentum space in the form,
 \bq
\lb{6.41}
u_{k}'' + \omega_{k}^{2}u_{k} = 0,
\eq
 where
\bqn
\lb{6.42}
\omega_{k}^{2} &=&  k^2{\cal{F}}_k - \frac{1}{4}\left({2{\cal{P}}' + {\cal{P}}^2  - 4\cal{Q}} \right), \nb\\
{\cal{F}}_k &\equiv& \frac{1}{\beta_0}\Bigg\{1+ 2 V_1 + 2\bar{A}({c_1}'-{c_2}) - 4 \pi G {c_1}^2  \left(1-\bar{A}\right)  \nb\\
&&  ~~~~~~~  +\frac{2  k^2 }{a^2} \left(V_2 +V'_4 + 2 \pi G \alpha_1 {c_1}^2 \right) \nb\\
&&  ~~~~~~~  -\frac{2  k^4 }{a^4} \left(V_6 + 2 \pi G \alpha_2 {c_1}^2\right)\Bigg\}.
\eqn
Note that the above hold only for $\lambda \not= 1$. When $\lambda =1$, we have a first-order equation for $\delta \chi$
\bq
\lb{6.43}
\delta \chi'  + \frac{\bar{\chi}'}{c_1} \left(c'_1 - f\right) \delta \chi =0,
\eq
which has the general solution,
\bq
\lb{6.44}
\delta \chi= \exp\left\{\int{\frac{\bar{\chi}'}{c_{1}}\big(f - c_1'\big)d\eta}\right\} \; \delta\chi_{1}(x),\; (\lambda = 1),
\eq
where $\delta\chi_{1}(x)$ is an arbitrary  function of $x$ only. Since in this paper we are mainly interested in the case $\lambda \not=1$, in the following
we shall not consider this case further.

Also, for the field to be stable in the UV regime, the condition 
\bq
\lb{6.44a}
V_6 + 2\pi G \alpha_2 {c_1}^2 <0,
\eq
 has to be satisfied. To study  Eq.(\ref{6.41}) further, we consider the sub- and super-horizon scales, separately. 

\subsection{Sub-Horizon Scales}

In this region, we have $k \gg {\cal{H}}$, and the dispersion relation reduces to, 
\bq
\lb{6.50}
\omega^2_k \simeq - \frac{2k^6}{\beta_0 a^4} \left(V_6+2\pi G \alpha_2 {c_1}^2\right).
\eq
With the extreme slow-roll condition,   we have $a\simeq -\frac{1}{H \eta}$ and $H, \; V_6, \; c_1\; \simeq$ Constants. Then, from  Eq. (\ref{6.41}) we find that
\bq
\lb{6.51}
u_k \propto e^{i \omega_k \eta}.
\eq

Unlike the case without the U(1) symmetry  \cite{WWM},  
 the 
metric perturbations $\psi$ and $B$ now oscillate with the same frequency as $\delta \chi$,  as one can see from Eqs.(\ref{6.17c}) 
and (\ref{6.17e}). Therefore, they are always coupled to the scalar field modes.

 \subsection{Super-Horizon Scales}
 
 In this region, we have   $k \ll {\cal{H}}$, and to the order of $k^2$, we find that
\bqn
\lb{6.60}
\omega^2_k \simeq&&   \frac{k^2}{\beta_0}\Big[1+ 2 V_1 + 2\bar{A}({c_1}'-{c_2}) - 4 \pi G {c_1}^2\left(1-\bar{A}\right) \Big]  \nb\\
&& +{\cal{Q}} -\frac{2{\cal{P}}' + {\cal{P}}^2}{4}.
\eqn
In the extreme slow-roll and massless limit ($\bar{\chi}' \simeq 0 \simeq V', V''\simeq0$), we obtain the following solution 
\bqn
\lb{6.61}
u_k &=& - \frac{D_1}{H\eta}\Bigg\{1+\frac{k^2 \eta^2}{2 \beta_0}\Big[1+ 2 V_1 \nb\\
&& ~~~ + 2\bar{A}({c_1}'-{c_2}) - 4 \pi G {c_1}^2 \left(1-\bar{A}\right) \Big]\Bigg\}\nb\\
&& ~~~ + D_2 \eta^2\Bigg\{1-\frac{k^2 \eta^2}{10 \beta_0}\Big[1+ 2 V_1 \nb\\
&& ~~~ + 2\bar{A}({c_1}'-{c_2}) - 4 \pi G  {c_1}^2 \left(1-\bar{A}\right) \Big]\Bigg\}\nb\\
&\sim& D_1 a + D_2 \eta^2,
\eqn
where the first term represents a constant perturbation, while the    second term represents a decaying mode.
Then, we find that
\bqn
\lb{6.62}
&&\delta \chi \simeq D_1 - D_2 H \eta^3, \nb\\
&&  \psi \simeq  4\pi G c_1\delta \chi, \nb\\
&& k^2 B  \simeq  -  \frac{12\pi G c_1}{|c_\psi|^2} D_2 \eta^2.
\eqn
In terms of the gauge-invariant quantities (\ref{4.0db}),    we obtain
\bqn
\lb{6.63}
&& \Psi_k = \psi_k - {\cal{H}}B_k, \nb\\
&&  \Phi_k = {\cal{H}}B_k + B'_k,\nb\\
&& \Phi_k-\Psi_k = H^2 k^2\eta^2 \left(H^2 k^2\eta^2 \alpha_2 - \alpha_1\right)\psi_k \nb\\
&& ~~~~ + H\eta\left(\hat{A}\psi_k - \delta A_k\right).
\eqn
Thus, like in the case without the U(1) symmetry \cite{WWM},   the dynamical evolution now leads to $\Phi=\Psi\rightarrow0$ at late times ($\eta\rightarrow0$).

\section{ Power Spectra and indices of Scalar and Tensor Perturbations}  

\renewcommand{\theequation}{6.\arabic{equation}} \setcounter{equation}{0}

 To calculate the spectra and indices of scalar and tensor  perturbations with the slow-roll approximations, we shall use the uniform approximation, 
 proposed recently in \cite{Hab}, and applied to the studies of
  tensor perturbations in the HL theory without the U(1) symmetry  in \cite{YKN,Wang}. We shall closely follow the treatment presented in
  \cite{Wang}. In particular,  for perturbations given by,
 \bq
 \lb{6.71a}
  v_k'' = [ g(k, \eta)+q(\eta)]v_k,
  \eq
  where $q(\eta) = -  1/4\eta^2$, and $v_k$ is the canonically normalized field,   the corresponding   power spectrum and  index  
  {at leading order of the uniform approximation} are given as  \cite{Wang}, 
  \bqn
   \lb{6.71b}
 \left. P_{v}(k)\right|_{k\eta  \rightarrow 0^{-}} &\equiv&  \left. \frac{k^{3}}{2\pi^{2}}\left|v_k\right|^{2}  \right|_{k\eta \rightarrow 0^{-}}\nb\\
 &=&  \lim_{k\eta \rightarrow 0^{-}}\frac{k^{3}  \exp\Big\{2{\cal{D}}(k, \eta)\Big\}}{4\pi^{2}a^{2}\sqrt{g(k, \eta)}},\\
\lb{6.71c}
 n_{v} - 1 &\equiv&  \left. \frac{d\ln{P_{v}}}{d\ln{k}} \right|_{k\eta \rightarrow 0^{-}},  
 \eqn
where
 \bq
 \lb{6.71d}
 {\cal{D}}({k,\eta}) \equiv \int^{\eta}_{\bar{\eta}(k)}{{\sqrt{g(k, \eta')}{d\eta'}}},
 \eq
 and $\bar{\eta}(k)$ denotes the turning point $g(k, \bar{\eta}) = 0$.  Note that in writing the above expressions, we assumed that there is only one turning point, that is,
 we consider only the case where $g(k,\eta) = 0$  
 has only one real root. For detail, see \cite{Wang}. In the following,  we shall apply the above  to the cases of scalar and tensor perturbations.
  
 \subsection{ Power Spectrum and Index of Scalar Perturbations}

 With the help of the master equation (\ref{6.38}) and the definition of gauge-invariant ${\cal{R}}$ in (\ref{6.6}), {the second order action reads},
\bqn
\lb{6.72a}
&& S^{(2)} = \frac{1}{2}\int d\eta d^3 x a^2 h^2\Big[\beta_0 {\cal{R}}'^2 -\beta_4 {\cal{R}}^2 -  \beta_1 (\partial_i{\cal{R}})^2\nb\\
&& ~~~~~~~~~~~~~~~~~~~~ -  \beta_2 (\partial^2 {\cal{R}})^2 - \beta_3 (\partial_i\partial^2 {\cal{R}})^2 \Big],
\eqn
where 
\bqn
\lb{6.72b}
\beta_0 &=& f +4 \pi G c^2_1 / |c^2_{\psi}|,\nb\\
%&=& f + \frac{1}{2}\left(\frac{c_1}{\Lambda_{\omega}}\right)^{2}\left(\frac{M_{\text{pl}}}{\Lambda_{\omega}}\right)^{2}, \nb\\
\beta_1 &\equiv&  \left[1+ 2 V_1 + 2 \bar{A} (c_1' - c_2) - 4\pi G c^2_1 (1-\bar{A})\right], \nb\\
\beta_2  &\equiv& \frac{2}{a^2} \left(V_2 + V_4' + 2\pi G c^2_1 \alpha_1\right), \nb\\
\beta_3 &\equiv& - \frac{2}{a^4} \left(V_6 + 2\pi G c^2_1 \alpha_2\right), \nb\\
\beta_4 &\equiv& \beta_0 {\cal{Q}} - \beta_0 \frac{h'^2}{h^2} + \frac{\left(a^2\beta_0hh'\right)'}{a^2 h^2}\nb\\
h &\equiv& \left(4\pi G c_1 + \frac{H}{\dot{\bar{\chi}}}\right)^{-1} = \frac{\delta \chi}{{\cal{R}}}. 
\eqn
After introducing the variable
\bq
\lb{6.72c}
v \equiv  z {\cal{R}}, \;\;\;
z^2 \equiv a^2 h^2 \beta_0,
\eq
the action is normalized to
\bqn
\lb{6.72d}
S^{(2)} &=& \frac{1}{2}\int d\eta d^3 x \left[ (v')^2 - \frac{\beta_1}{\beta_0} (\partial_i v)^2 - m^2_{\text{eff}}v^2\right] \nb\\
 && -\frac{1}{2} \int d\eta d^3 x \left[ \frac{\beta_2}{\beta_0} (\partial^2 v)^2 + \frac{\beta_3}{\beta_0}(\partial_i\partial^2 v)^2 \right] . \nb\\
\eqn
{Here $m^2_{\text{eff}}$ is defined to be}
\bqn
\lb{6.72e}
- m^2_{\text{eff}} \equiv \frac{z''}{z} - \frac{\beta_4}{\beta_0} .
\eqn
Going through the quantization procedure as described in Appendix C, the classical equation of motion for mode functions 
$v_k$ are
\bq
\lb{6.72f}
v''_k + \left(\omega^2_k  + m^2_{\text{eff}}\right) v_k =0, 
\eq
where
\bq
\lb{6.72fa}
\omega^2_k  = \frac{k^2}{\beta_0} \left(\beta_1 + {\beta_2} k^2 + {\beta_3} k^4\right).
\eq

 Looking at the expressions of the $\beta$  coefficients, we see that they contain terms of $c_1, c_2, V_1, V_2, V_4, V_6$ and $\bar{A}$. Now go back to the Lagrangian describing the inflaton $\chi$,  Eqs.(\ref{5.2}) and (\ref{5.7}), one can see that $V_1, V_2, V_4, V_6$ all stem from the potential term ${\cal{V}}$, while $c_1$ and $c_2$ appear through the first line of (\ref{5.7}), which  can also be taken as a ``potential term''. (Note that the second and the third line of (\ref{5.7}) correspond to modifications of dynamical coupling terms due to the presence of $\varphi$.).  Therefore, we could assign their respective ``slow-roll" parameters describing their time evolution during inflation in a manner similar to $V$. However, unlike $V$, which appears in the background equation (\ref{5.13a}), these ``potential terms" are not constrained by the background equations. As an approximation, we assume here that the time dependence of $c_1, c_2, V_1, V_2, V_4, V_6$ and $\bar{A}$ are at least second order in terms of the slow roll parameters. Since we only consider the first order approximations in this paper, they can be taken as constant throughout inflation. This also leaves $d \beta_0/d\eta \propto d c_1/ d\eta=0$.
 
 With the above assumptions, it can be shown  that $h^2$ relating $\delta \chi$ and ${\cal{R}}$ is of order ${\cal{O}} (\epsilon_{{\scriptscriptstyle V}})$. In fact, from its definition,
 \bqn
 \lb{6.73a}
 h^2 &=&\left(4\pi G c_1 + \frac{H}{\dot{\bar{\chi}}}\right)^{-2} = \left(\frac{\dot{\bar{\chi}}}{H}\right)^2 \left[1+ \frac{c_1 \dot{\bar{\chi}}}{2 M^2_{\text{pl}} H}\right]^{-2}\nb\\
 &=& 2 \tilde{M}^2_{\text{pl}} \epsilon_{{\scriptscriptstyle V}} \left[1+ \frac{c_1 \dot{\bar{\chi}}}{2 M^2_{\text{pl}} H}\right]^{-2}.
 \eqn
 Since 
 \bq
 \lb{c1}
\frac{ c_1 \dot{\bar{\chi}}}{ 2M^2_{\text{pl}} H} = \frac{c_{1}}{\sqrt{2}M_{\text{pl}}}\times\frac{1}{\sqrt{2}M_{\text{pl}}}\frac{\dot{\bar{\chi}}}{H} =\frac{c_{1}}{\sqrt{2}M_{\text{pl}}} \sqrt{\tilde{\epsilon}_{{\scriptscriptstyle V}}} \ll 1,\nb\\
 \eq
  where $|c_{1}| \simeq M_{*} \ll M_{\text{pl}}$ \cite{LWWZ}, we have, to first order of the slow-roll parameters, 
 \bq
  \lb{6.73b}
 h^2 \simeq 2 \tilde{M}^2_{\text{pl}} \epsilon_{{\scriptscriptstyle V}} .
 \eq
  {On the other hand, $a(\eta) \simeq -(1+\epsilon_{{\scriptscriptstyle V}})/(H\eta)$, which  leads to %the mass term
  \bq
  \lb{6.74}
  -m^2_{\text{eff}} \simeq \frac{2-3\eta{{\scriptscriptstyle V}} + 9 \epsilon{{\scriptscriptstyle V}}}{\eta^2} + \frac{\Delta m^2}{\eta^2}.
  \eq
Here the first term comes from $z''/z$ and is the same as that from GR under the above assumptions, whereas the second term introduces new effects,
  \bqn
  \lb{6.74b}
  \Delta m^2 &\simeq& \frac{1}{\beta_0}\Bigg[3\left(\beta_0-1\right)\eta{{\scriptscriptstyle V}}+\left(\frac{2f^2}{\lambda-1}-6\beta_0\right) \epsilon_{{\scriptscriptstyle V}} \nb\\
  			&&~~~~~ -\frac{2}{\lambda-1}\left(\frac{c_{1}}{\sqrt{2}M_{\text{pl}}} \sqrt{\epsilon_{{\scriptscriptstyle V}}}\right)\Bigg].
  \eqn
  We see that the these are in general functions of $\lambda$. Since it's the time-dependence of $m_{\text{eff}}$ which breaks the exact scale-invariance, we would expect that observations on the power index, which will be derived below, place constraints on the value of $\lambda$.}
  
  The function $g(k, \eta)$ defined through Eqs.(\ref{6.71a}) and (\ref{6.72f})  is now given by
  \bq
  \lb{6.75a}
  g(k, \eta) = \frac{k^2}{y^2} \left[a^2_0 -\left(a_2 y^2 + a_4 y^4 + a_6 y^6 \right)\right],
  \eq
  where
  \bqn
  \lb{6.75b}
  y &\equiv& k\eta, \nb\\
  a^2_0 &\equiv& \frac{1}{4} - m^2_{\text{eff}} \eta^2 = \frac{9}{4} -3\eta{{\scriptscriptstyle V}} + 9 \epsilon{{\scriptscriptstyle V}}+\Delta m^2, \nb\\
  a_2 &\equiv& \frac{1}{\beta_0}\left[1+2V_1+ 2 \bar{A} (c_1' + c_2) - 4 \pi G c^2_1 (1-\bar{A})\right], \nb\\
  a_4 &\equiv& \frac{1}{\beta_0}\left[2H^2 (1-2\epsilon_{{\scriptscriptstyle V}}) \left(V_2 + V_4' + 2\pi G c^2_1 \alpha_1\right)\right],\nb\\
  a_6 &\equiv& - \frac{1}{\beta_0}\left[2H^4(1-4\epsilon_{{\scriptscriptstyle V}}) \left(V_6+ 2\pi G c^2_1 \alpha_2\right)\right].
  \eqn
  Thus the power spectrum  of ${\cal{R}}$ is given by,
  \bqn
  \lb{6.76a}
    P_{{\cal{R}}}(k) &=&  \frac{H^2 (1-2\epsilon_{{\scriptscriptstyle V}}) |y_0|^3}{4\pi^2 \beta_0 a_0 h^2} \left(\frac{2}{e}\right)^{2a_0}  \text{exp} \left[\frac{a_4 y^4_0 + a_6 y^6_0}{3a_0}\right]\nb\\
  && \times \lim_{y \rightarrow 0} \left(\frac{y}{y_0}\right)^{3-2a_0},
  \eqn
   where $y_0$ is defined to be the turning point of $g(k, \bar{\eta})$, namely,  
  \bq
  \lb{6.77}
   a_6 y_0^6   + a_4 y_0^4 +   a_2 y_0^2 - a^2_0   = 0.
  \eq
  Clearly,   $y_0$ is independent  of $k$, as $a_{n}$'s are. {Then,  we find
  \bqn
   \lb{6.76b}
  n_{{\scriptscriptstyle {\cal{R}}}} - 1 &=& 2\eta_{{\scriptscriptstyle V}} - 6 \epsilon_{{\scriptscriptstyle V}} - \frac{2}{3} \Delta m^2.
%  \nb\\
%  				&& +\frac{k}{y_0} \frac{dy_0}{dk}\Bigg[3+6\epsilon_{{\scriptscriptstyle V}}-2\eta_{{\scriptscriptstyle V}} + \frac{2}{3}\Delta m^2 \nb\\
%  			&&   ~~~~~~~~~~~~ + \frac{2}{9}\left(1-2\epsilon_{{\scriptscriptstyle V}}+\frac{2}{3}\eta_{{\scriptscriptstyle V}}+ \frac{2}{9}\Delta m^2\right)\nb\\
%   && ~~~~~~~~~~~~~~~ \times \left(4a_4 y^4_0 + 6a_6 y^6_0\right)\Bigg]\nb\\
%   &=& 2\eta_{{\scriptscriptstyle V}} - 6 \epsilon_{{\scriptscriptstyle V}} - \frac{2}{3} \Delta m^2\nb\\
%   &=& \frac{2}{\beta_0} \eta_{{\scriptscriptstyle V}} \nb\\
%   && - \Big[2+ \frac{4}{3\beta_0(\lambda-1)}\left(f^2-\frac{c_1}{\sqrt{2}M_{\text{pl}} \sqrt{\epsilon_{{\scriptscriptstyle V}}}}\right)\Big]\epsilon_{{\scriptscriptstyle V}}.\nb\\
  \eqn
  Due to correction term $\Delta m^2$, the spectrum index of the scalar perturbation does not reproduce the GR value in general. In particular, we see that it  depends on  the value $1/(\lambda-1)$ (Note that $\beta_0$ is also a function of $\lambda-1$). One may worry that in the relativistic limit at low energy $\lambda \rightarrow 1$ these terms will diverge, hence breaking the near-scale-invariance of the spectrum. However, during inflation, we are in a region where UV physics dominates, thus the value of $\lambda$ is expected far away from its relativistic fixed point at that time.  }%textcolor{red}
  
   {We note that even when the index can restore the standard value in GR,\footnote{Mathematically, in the limit $c_1=0, \beta_0 \rightarrow 1$ and $f^2/3\beta_0(\lambda-1) \rightarrow 1$, the standard result can be restored.} it is a consequence of our assumption that all the potential terms $c_1, c_2, V_1, V_2, V_4$ and $V_6$ are time-independent. Comparing the definition of $z$ given here  in Eq.(\ref{6.72c})  with the one given in GR, we can see some extra terms of $c_1$ appear. By our assumption, $d c_1/d\eta=0$, this makes $d\beta_0/d\eta=0$ and $d h^2/d\eta \propto d (\dot{\bar{\chi}}/H)/d\eta$, which leaves the term $z''/z$ exactly the same as that of a single field in GR. Further more, in the modified dispersion relation (\ref{6.72fa}),  the term $\beta_1/\beta_0$ corresponds to the relativistic case,  and the ones $\beta_2/\beta_0$ and $ \beta_3/\beta_0$ are induced by Lorentz-symmetry-breaking effects, which are  assumed to be time-independent. If,  however, these potential terms are evolving with time during inflation, one needs to take into account of the time-dependence of the dispersion relation and of the varying effective mass \cite{Martin}. The  same arguments also apply to the studies of tensor spectrum and index.}%textcolor{red}
  
{We would also like to note that, the exact form of scale-dependence of the scalar spectrum depends on the instant when it's evaluated \cite{Stew}, and can receive further corrections when we incorporate a second order uniform approximation \cite{Hab}. What's more, from an observational point of view, as long as the scale-dependence is not broken severely, the connection between tensor-scalar-ratio and slow-roll parameters is more important than the tilt itself.}%textcrolor{red}

  Setting the slow roll parameters to zero  exactly, the power spectrum given above can be put in the simple form, 
  \bqn
  \lb{6.76c}
    P_{{\cal{R}}}(k) &=& \frac{4 H^4 |y_0|^3}{3\pi^2 e^3 \beta_0 \dot{\bar{\chi}}^2}\nb\\ % \left(1+ \frac{c_1 \dot{\bar{\chi}}}{2 M^2_{\text{pl}}H}\right)^2 \nb\\
  && \times \; \text{exp}\left[\frac{2}{9} \left(a_4 y^4_0+ a_6 y^6_0\right)\right].
  \eqn
  In the relativistic limit ($a_2=\beta_0=1, a_4=a_6=c_1=0$, and $g_{s} = 0, (s = 2, ..., 8)$, this yields the well-known result obtained in GR \cite{Inflation},
  \bqn
  \lb{6.91}
  P^{GR}_{{\cal{R}}} &=& \frac{18}{e^3} \Big(\frac{H^2}{2\pi\dot{\bar{\chi}}}\Big)^2,
  \eqn
 except for the factor   $ 18/e^3 \sim 0.896$. This difference in magnitude is due to the way we normalize 
the power spectrum in the uniform approximations.
As shown later, the same factor also appears in the expression for the power spectrum of tensor perturbations, so that the ratio of them does not depend on this factor. 

%    In the limit where the higher curvature terms's contributions are negligible, we have $y_0 \rightarrow -3/2 - 3\epsilon_{{\scriptscriptstyle V}} + \eta_{{\scriptscriptstyle V}}$, leading to 
%  \bqn
%  \lb{6.77a}
%  P_{{\cal{R}}}(k) &=& \frac{18}{e^3} \Big(\frac{H^2}{2\pi\dot{\bar{\chi}}}\Big)^2 , \\
%    \lb{6.77}
%  n_{{\scriptscriptstyle {\cal{R}}}} - 1 &=& -2 \eta_{{\scriptscriptstyle V}} + 6 \epsilon_{{\scriptscriptstyle V}}.
%  \eqn
%It is remarkable to note that the tilt is exactly the same as that given in GR \cite{Inflation}.

 To estimate  the effect from higher curvature terms on the power spectrum, let us first  write the dispersion relation (\ref{6.72fa}) in the  form,
 \bq
\lb{6.83}
\omega^2_k \equiv b_1 k^2 + b_2 \frac{k^4}{a^2 M^2_{\text{A}}} + b_3\frac{k^6}{a^4 M^4_{\text{B}}},
\eq
where   \cite{HW,LWWZ}
\bq
\lb{6.82}
 M_{\text{A}} \equiv |g_{3}|^{-1/2} M_{\text{pl}}, ~~~~~ M_{\text{B}} \equiv |g_{8}|^{-1/4} M_{\text{pl}},
\eq
and
\bqn
\lb{6.83a}
b_1 &\equiv& \left[1+2V_1+ 2 \bar{A} (c_1' + c_2) - 4 \pi G c^2_1 (1-\bar{A})\right]/\beta_0,\nb\\
b_2 &\equiv& 2\left[\lambda_2 + \lambda_4 + \lambda_{23}\right]/\beta_0,\nb\\
b_3 &\equiv&  -2\left[\lambda_6+\lambda_{78}\right]/\beta_0,
\eqn
with
\bqn
\lb{6.84}
&& \lambda_2 \equiv V_2 M^2_{\text{A}}, ~~~ \lambda_4 \equiv V'_4 M^2_{\text{A}}, ~~~ \lambda_6 \equiv V_6 M^4_{\text{B}}, \nb\\
&& \lambda_{23} \equiv \frac{c^2_1 M^2_{\text{A}}}{2 M^4_{\text{pl}}}(8g_2+3g_3),\nb\\
&& \lambda_{78} \equiv \frac{c^2_1 M^4_{\text{B}}}{   M^6_{\text{pl}}}(8g_7- 3g_8).
\eqn
%\bqn
%\lb{6.80}
%\omega^2_k = \mu_1 k^2+ \mu_2 \frac{k^4}{a^2M^2_A} +  \mu_3 \frac{k^6}{a^4M^4_B},
%\eqn
Since $g_{2}$ and $g_{3}$ all both the coefficients of the fourth-order derivative terms, as one can see from Eq.(\ref{2.5a}), it is quite reasonable to assume that
$g_{2}$ and $g_{3}$ are in the same order, $g_{3}/g_{2} \simeq {\cal{O}}(1)$. Similarly, one can argue that
$g_{s}/g_{4}  \simeq {\cal{O}}(1)$ for $s = 5, 6, 7, 8$, as all of these terms are the coefficients of the sixth-order derivative terms. For the sake of simplicity, we further
assume $M_{\text{A}} \simeq M_{\text{B}} = M_{*}$. Taking $c_1 \simeq c_2 \simeq M_{*}$, from Eq.(\ref{SC}) we find $M_{*} \le  M_{\text{pl}}|c_{\psi}|^{1/2}$ \cite{LWWZ}.
Then, from Eq.(\ref{6.72b}) we obtain
\bq
\lb{6.72ba}
\beta_0 \simeq  f(\lambda) + \frac{1}{2}\left(\frac{M_*}{\Lambda_{\omega}}\right)^{4/5} \simeq  {\cal{O}}(1), %\left(\frac{M_{\text{pl}}}{\Lambda_{\omega}}\right)^{2},
\eq
as $f(\lambda) \simeq {\cal{O}}(1)$. 
%
%Thus, for $\beta_0 \simeq {\cal{O}}(1)$, we require
%\bq
%\lb{M*}
%M_* \leqslant \left(\frac{\Lambda_{\omega}}{M_{\text{pl}}}\right)\Lambda_{\omega}.
%\eq
To determine the scales of $V_{n}$, we  assume that $b_{n}$ defined in Eq.(\ref{6.83a})  are all of order 1,
i.e., 
\bq
\lb{bs}
b_{n} \simeq  {\cal{O}}(1), (n = 1, 2, 3),
\eq
which is a reasonable assumption, considering the physical meanings of the energy scales $M_{\text{A}}$ and $M_{\text{B}}$. In fact, one can define $M_{\text{A}}$ and $M_{\text{B}}$ so that
$b_{2} = b_{3} = 1 $ precisely, as originally defined in    \cite{LWWZ}. To have $b_1 = 1$, one can properly choose $V_1$. On the other hand, since $\bar{A}$ is undetermined, and for the sake
of simplicity, we further set $\bar{A} = 0$. %the background 

%. For scalar perturbation ${\cal{R}}$, the dispersion relation reads
%\bqn
%\lb{6.81}
%\omega^2_k &=& \frac{1}{\delta_1}\Bigg\{1+ 2 V_1 + 2\bar{A}({c_1}'-{c_2}) - 4 \pi G {c_1}^2  \left(1-\bar{A}\right)  \nb\\
%&& +\frac{2  k^2 }{a^2} \left(V_2 +V'_4 + 2 \pi G \alpha_1 {c_1}^2 \right) \nb\\
%&& -\frac{2  k^4 }{a^4} \left(V_6 + 2 \pi G \alpha_2 {c_1}^2\right)\Bigg\}.
%\eqn
%Now define (the specific form is for later convenience with tensor perturbations.)

%then the dispersion becomes \footnote{Here we put slow roll parameters to zero since we are only dealing with power spectrum.}
%where we have introduced dimensionless parameters 

%\textcolor{red}{For good physics, these parameters should, in general, be in the order of ${\cal{O}}(1)$. }Accordingly,

With all the above assumptions, we find that  the function $g(k, \eta)$ now reads,
\bqn
\lb{6.85}
g(k, \eta) &=& \frac{k^2}{y^2}\left[\frac{9}{4}-y^2 \left(1 +   \epsilon_{{\scriptscriptstyle\text{HL}}} y^2 +  \epsilon_{{\scriptscriptstyle \text{HL}}}^2 y^4\right)\right], \nb\\
\epsilon_{{\scriptscriptstyle \text{HL}}} &\equiv& \frac{H^2}{M^2_*}.
\eqn
Thus, depending on the energy scale $H$ when inflation occurs, one can have different turning point $y_0$. In the following we consider only  two limits,
$\epsilon_{{\scriptscriptstyle \text{HL}}} \ll 1$ and $\epsilon_{{\scriptscriptstyle \text{HL}}} \gg 1$. In addition, in writing Eq.(\ref{6.85}) we have
set $b_{n} = 1  = \beta_{0}$ precisely.  General expressions without setting $b_n  =1$ can be found in Appendix D.

\subsubsection{$\epsilon_{{\scriptscriptstyle \text{HL}}} \ll 1$}

When $\epsilon_{{\scriptscriptstyle \text{HL}}} \ll 1$, to its second order, we find that 
% If we treat $H^2/M^2_A$ and $H^2/M^2_B$ as small quantities, we can get the perturbative solution of $g(y_0)=0$. Assuming that $M_A\simeq M_B = M_\star$
\bqn
\lb{6.86}
y^2_0 \simeq \frac{9}{4} \left(1- \frac{9}{4} \epsilon_{{\scriptscriptstyle \text{HL}}}  + \frac{81}{16} \epsilon_{{\scriptscriptstyle \text{HL}}}^2\right), 
%\frac{H^2}{M^2_\star} + \frac{81(2b^2_2-b_1b_3)}{32b^4_1}\frac{H^4}{M^4_\star}\right].
\eqn
for which  the power spectrum  is given by, 
\bqn
\lb{6.90}
P_{{\cal{R}}}(k)  &\simeq&  {P^{\text{GR}}_{{\cal{R}}}} \left(1- \frac{9}{4} \epsilon_{{\scriptscriptstyle \text{HL}}}  + \frac{729}{128} \epsilon_{{\scriptscriptstyle \text{HL}}}^2\right).\nb\\
%\left(1+\frac{c_1\dot{\bar{\chi}}}{2M^2_{\text{pl}}H}\right)^2\nb\\
% &\times& \Bigg\{1-\frac{9b_2}{4b^2_1}\frac{H^2}{M^2_\star} + \frac{81(17b^2_2-8b_1b_3)}{128b^4_1}\frac{H^4}{M^4_\star}\Bigg\}. \nb\\
\eqn

It is interesting to note that the condition $\epsilon_{{\scriptscriptstyle \text{HL}}} \ll 1$ is equivalent to 
%\textcolor{red}{The assumption that $H^2/M^2_\star$ is small has a relation with inflation model building.} From the Friedmann equation (\ref{5.13a}), during inflation,
\bq
\lb{6.95}
  V(\bar\chi) \ll \frac{3}{2}(3\lambda-1) \left(\frac{\Lambda_{\omega}}{M_{\text{pl}}}\right)^{2} M^4_{\text{pl}},  ~~~ (\epsilon_{{\scriptscriptstyle \text{HL}}} \ll 1).
\eq

\subsubsection{$\epsilon_{{\scriptscriptstyle \text{HL}}} \gg 1$}

When $\epsilon_{{\scriptscriptstyle \text{HL}}} \gg 1$, to find the turning point $y_{0}$, we first write $g(k, \eta)$ given by Eq.(\ref{6.85}) in the form,
\bqn
\lb{6.85r}
g(k, \eta) = \epsilon^2_{{\scriptscriptstyle \text{HL}}}\frac{k^{2}}{y^2}\Big[\frac{9}{4}\eta_{{\scriptscriptstyle \text{HL}}}^{2} 
-y^2 \left(\eta_{{\scriptscriptstyle \text{HL}}}^{2} + \eta_{{\scriptscriptstyle \text{HL}}} y^2 +  y^4\right)\Big],\;\;
\eqn
where $\eta_{{\scriptscriptstyle \text{HL}}} \equiv 1/\epsilon_{{\scriptscriptstyle \text{HL}}} \ll 1$. Then we find the perturbative solution
%from Eq.(\ref{6.85}) we find that $y_{0}$ is simply given by,
\bq
\lb{6.95a}
y^2_{0} \simeq \left(\frac{3\eta_{{\scriptscriptstyle \text{HL}}}}{2}\right)^{2/3}\left[1 - \frac{1}{3}\left(\frac{4\eta_{{\scriptscriptstyle \text{HL}}}}{9}\right)^{1/3}
-\frac{2}{9}\left(\frac{4\eta_{{\scriptscriptstyle \text{HL}}}}{9}\right)^{2/3}\right],
\eq
for which  the power spectrum takes the form, 
\bqn
\lb{6.95b}
P_{{\cal{R}}} (k) &\simeq&  {P^{\text{GR}}_{{\cal{R}}}(k)} \frac{4 \eta_{{\scriptscriptstyle \text{HL}}} \sqrt{e}}{9} \Bigg[1- \frac{1}{2}\left(\frac{4\eta_{{\scriptscriptstyle \text{HL}}}}{9}\right)^{1/3}\Bigg].\nb\\
\eqn
Thus, if the inflation    happened way above the scale $M_{*}$, the spectrum will be suppressed by the factor $M^2_{*}/H^2$, comparing with that of GR.

%$\lambda$ is usually limited to be close to 1 due to the IR behavior of the theory, the power spectrum normalization from experiments 
%further constrains the magnitude of power spectrum. These two combined, it might indicate that inflation in HL theory happens at a much lower 
%scale, comparing to the case of GR.

 \subsection{ Power Spectrum and Index of Tensor Perturbations}
 
The tensor perturbations can be written in the form \cite{WM,Wang}
\bq
\lb{6.100}
\delta g_{ij} = a^2 \left( \delta_{ij} + h_{ij}\right),
\eq
where $h_{ij}$ is traceless and transverse, i.e., $h^i_{\;\; i} =0=\partial^{j} h_{ij}$. For a single scalar inflaton, the anisotropic stress is zero, so the  tensor perturbations are
 source-free.
In the ADM formalism, with the results of constraint equations derived in Section II, it can be shown that  the second order action is given by
%\textcolor{red}{(check if any slow-roll approximations are made here)}
\bqn
\lb{6.101}
S^{(2)} &=& \frac{1}{2} \int d\eta d^3 x \frac{\zeta^2 a^2}{2} \Bigg\{\left(\partial_\eta h_{ij}\right)^2 - (1-\bar{A}) \left(\partial_c h_{ij}\right)^2 \nb\\
&& ~~ - \frac{g_3}{\zeta^2 a^2}\left(\partial^2 h_{ij}\right)^2 - \frac{g_8}{\zeta^4 a^4}\left(\partial_c \partial^2 h_{ij}\right)^2\Bigg\}.
\eqn
 Defining the following expansion in the momentum space \cite{Inflation},
\bq
\lb{6.102}
h_{ij} = \int \frac{d^3 k}{(2 \pi)^3} \sum_{s=+,\times} \epsilon^s_{ij}(k) h^s_\mathbf{k}(\eta) e^{i\mathbf{k}\mathbf{x}},
\eq
where $\epsilon_{ii} = k^i \epsilon_{ij} =0$ and $\epsilon^s_{ij}(k) \epsilon^{s'}_{ij}(k)=2\delta_{ss'}$, the above action becomes
\bqn
\lb{6.103}
S^{(2)} &=& \sum_{s=+, \times} \int d\eta d^3 k\frac{a^2}{2\zeta^2}\Bigg\{(h^{'s}_\mathbf{k})^2- (1-\bar{A}) k^2 (h^s_\mathbf{k})^2\nb\\
 && - \frac{g_3 k^4}{\zeta^2 a^2} (h^s_\mathbf{k})^2 - \frac{g_8 k^6}{\zeta^4 a^4} (h^s_\mathbf{k})^2\Bigg\}.
\eqn
To make the action canonically normalized, we introduce $v^s_\mathbf{k}$ by
\bq
\lb{6.105}
v^s_\mathbf{k} \equiv a \zeta h^s_\mathbf{k}.
\eq
Then, the action (\ref{6.103}) becomes 
\bq
\lb{6.106}
S_{(2)} = \frac{1}{2}\sum_{s=+, \times} \int d\eta d^3 k \Bigg[(v^{'s}_\mathbf{k})^2 - (\omega^2_k + m^2_{\text{eff}}) (v^{s}_\mathbf{k})^2\Bigg],
\eq
but now with 
 \bqn
 \lb{6.107}
 \omega^2_k &=& (1-\bar{A})k^2 + \frac{g_3 k^4}{\zeta^2 a^2} + \frac{g_8 k^6}{\zeta^4 a^4}, \nb\\
 m^2_{\text{eff}} &=& -\frac{a''}{a}.
 \eqn
 One can see that each spin state of the tensor perturbation acts like a scalar. After the quantization procedure prescribed in Appendix C, the classical equation of motion 
  for the mode functions again read
 \bq
 \lb{6.108}
v''_k + (\omega^2_k + m^2_{\text{eff}}) v_k =0,
 \eq
 where in writing the above equation, we had dropped the super indices ``s", and  $\omega^2_k$ and $ m^2_{\text{eff}}$ are now defined by Eq.(\ref{6.107}).
 From the above, we can directly read off $g(k, \eta)$ for tensor perturbations, 
    \bqn
   \lb{6.110}
   g(k,\eta) &=& \frac{k^2}{y^2}\Bigg\{\frac{9}{4}(1+2\epsilon_{{\scriptscriptstyle V}}) - \Big[(1-\bar{A})y^2\nb\\
    &&  +  \frac{g_3 H^2}{\zeta^2}(1-2\epsilon_{{\scriptscriptstyle V}}) y^4 
    + \frac{g_8 H^4}{\zeta^4} (1-4\epsilon_{{\scriptscriptstyle V}})y^6\Big]\Bigg\}, \nb\\
   y&\equiv&  k\eta.
   \eqn
Thus,    its turning point $y^2_0 \equiv (k \bar{\eta})^2$ satisfies the  cubic equation
\bqn
   \lb{6.111}
   \frac{9}{4} (1+2\epsilon_{{\scriptscriptstyle V}}) &=& (1-\bar{A})y^2_0 + \frac{g_3 H^2}{\zeta^2}(1-2\epsilon_{{\scriptscriptstyle V}}) y^4_0   \nb\\
   && ~~~~~~~~~~ + \frac{g_8 H^4}{\zeta^4} (1-4\epsilon_{{\scriptscriptstyle V}})y^6_0.
   \eqn 
Then the dimensionless spectrum and index for the tensor perturbations can be defined as \cite{Inflation}, 
 \bqn
 \lb{6.112}
\left. P_{\text{T}}(k)\right|_{k\eta  \rightarrow 0^{-}} &\equiv&  4 \times \left. \frac{k^{3}}{2\pi^{2}}\frac{\left|v_k\right|^{2}}{\zeta^2 a^2}\right|_{k\eta \rightarrow 0^{-}} ,\\
 \lb{6.113}
 n_{{\scriptscriptstyle \text{T}}} &\equiv& \left. \frac{d\ln{P^2_{\text{T}}}}{d\ln{k}} \right|_{k\eta \rightarrow 0^{-}}. 
 \eqn
 Here the factor of 4 accounts for the two spin states.  
 
Again, assuming that the gauge field $\bar{A}$ is constant during inflation, %\textcolor{red}{the power spectrum and index defined above are given by}
 \bqn
 \lb{6.114}
 P_{\text{T}}(k) &=& \frac{16 H^2 |y_0|^3}{3\pi^2 e^3 \zeta^2} \nb\\
 && \times \text{exp}\left[\frac{2 H^2 y^4_0}{9 \zeta^2} \left(g_3 + g_8 y^2_0 \frac{H^2}{\zeta^2} \right)\right],
 \\
  \lb{6.115}
n_{{\scriptscriptstyle \text{T}}}  &=& -2\epsilon_{{\scriptscriptstyle V}}.
  \eqn

%  When $g_3=0=g_8$, we find that $y_0 = k \bar{\eta} = -(3+2\epsilon_{{\scriptscriptstyle V}})/2$, and the above 
% reduce to the well-known results obtained in GR, except for the factor  $18/e^3 \sim 0.896$, appearing in the expression of  the power spectrum
% $P_{\text{T}}(k) $.

 In the relativistic limit, Eq.(\ref{6.114}) yields the well-known results obtained in GR \cite{Inflation}
 \bqn
 \lb{6.117}
 P^{\text{GR}}_{\text{T}}(k) = \frac{18}{e^3} \frac{2H^2}{\pi^2 M^2_{\text{pl}}}.
 \eqn
 Because of the normalization of the power spectrum in the uniform approximation, a difference of a factor
 $18/e^3$ also appears in the tensor perturbations. 
 
  To study the effect of high  order curvature terms, following what we did for the scalar perturbations, we consider the two cases
  $\epsilon_{{\scriptscriptstyle \text{HL}}} \ll 1$ and $\epsilon_{{\scriptscriptstyle \text{HL}}} \gg 1$, separately. 
  
  \subsubsection{$\epsilon_{{\scriptscriptstyle \text{HL}}} \ll 1$}
  
  In this case,    the power spectrum (\ref{6.114})  takes the form
 \bqn
 \lb{6.116}
 P_{\text{T}}(k) \simeq  P^{\text{GR}}_{\text{T}}(k) \left(1-\frac{9}{2}\epsilon_{{\scriptscriptstyle \text{HL}}} +\frac{729}{32}\epsilon_{{\scriptscriptstyle \text{HL}}}^2\right).
 \eqn
Then,   from Eqs.(\ref{6.90}) and (\ref{6.116}), we   find that the scalar-tensor ratio is given by
 \bqn
 \lb{6.118}
 r &\equiv& \frac{P_{\text{T}}(k)}{P_{{\cal{R}}}(k)} \simeq 16\epsilon_{{\scriptscriptstyle V}}\left(1 - \frac{9}{4}\epsilon_{{\scriptscriptstyle \text{HL}}} + \frac{2187}{128}\epsilon_{{\scriptscriptstyle \text{HL}}}^2\right).  
 %\epsilon_{{\scriptscriptstyle \text{HL}}} - ?? \epsilon_{{\scriptscriptstyle \text{HL}}}^2\right), \nb\\
% &=&  \frac{4\beta_1}{\zeta^2}\left(\frac{y_{0_{\text{T}}}}{y_{0_{\text{S}}}}\right)^3 \left(\frac{\dot{\bar{\chi}}}{H}\right)^2 \left[1+ \frac{c_1\dot{\bar{\chi}}}{4 \zeta^2 H}\right]^{-2}\nb\\
%  && \times\; \text{exp} \Bigg\{\frac{2}{9}  y^4_{0_{\text{T}}} \left(g_3\frac{H^2}{\zeta^2}+g_8 \frac{H^4}{\zeta^4} y^2_{0_{\text{S}}}\right)\Bigg\}\nb\\
%% && \times \;  \text{exp} \Big\{- \frac{2}{9} y^4_{0_{\text{S}}} \left(a_2 + a_3 y^2_{0_{\text{S}}}\right)\Big\},
 \eqn 
For the general case, see Eq.(\ref{C1.2}).
% One may find in Appenix C that the specific value of $16\epsilon_{{\scriptscriptstyle V}}$ for this $r$ is due to our choice of $b_1=1$. If, however, $b_1$ 
%resulted from $V_1$ and $c_1$ is not exactly 1, the ratio will be differnt. But it will remain the same order since $b_1 \simeq {\cal{O}}(1)$.
 
 \subsubsection{$\epsilon_{{\scriptscriptstyle \text{HL}}} \gg 1$}
 
 When $\epsilon_{{\scriptscriptstyle \text{HL}}} \gg 1$, from Eq.(\ref{6.110}) we find that $y^2_{0}$ is  given by,
\bqn
\lb{6.119}
y^2_{0} \simeq&& \left(\frac{3\eta_{{\scriptscriptstyle \text{HL}}}}{4}\right)^{2/3} \Bigg[1-\frac{1}{6}\left(\frac{16\eta_{{\scriptscriptstyle \text{HL}}}}{9}\right)^{1/3}\nb\\
&& ~~ -\frac{1}{18}\left(\frac{16\eta_{{\scriptscriptstyle \text{HL}}}}{9}\right)^{2/3}\Bigg],
\eqn
and the power spectrum takes the form, 
\bqn
\lb{6.120}
P_{\text{T}} (k) &\simeq&  P^{\text{GR}}_{\text{T}}(k) \frac{2e^{1/2} \eta_{{\scriptscriptstyle \text{HL}}}}{9} \Bigg[1- \frac{1}{4}\left(\frac{16\eta_{{\scriptscriptstyle \text{HL}}}}{9}\right)^{1/3}\Bigg].\nb\\
\eqn
Then, the combination of it with  Eq.(\ref{6.95b}) yields 
 \bqn
 \lb{6.121}
 r  &\simeq& 8\epsilon_{{\scriptscriptstyle V}} \Bigg[1+\frac{2-2^{2/3}}{4}\left(\frac{4\eta_{{\scriptscriptstyle \text{HL}}}}{9}\right)^{1/3}\Bigg].
 \eqn 
For arbitrary $b_n$, see Eq.(\ref{C2.2}).

\section{Conclusions}

\renewcommand{\theequation}{7.\arabic{equation}} \setcounter{equation}{0}

In this paper,  we   have studied inflation driven by a single scalar field in the HMT setup \cite{HMT} with the projectability condition
and an arbitrary coupling constant 
$\lambda$ \cite{Silva}. Because of the particular coupling of matter fields (\ref{coupling}), in Sec. III.A
we have been able to
show that the FRW universe is necessarily flat for (multi-) scalar, vector and fermionic fields. It is quite reasonable to argue that this should be true for
all the viable cosmological models \footnote{It should be
noted that this conclusion is based on the coupling of matter fields to the gauge field $A$ and Newtonian prepotential $\varphi$, proposed in \cite{Silva}.}. 
Therefore, the HMT setup  provides a built-in mechanism to solve the flatness problem. However, to solve
 the horizon problem inflation may or may not be needed \cite{WOinf}, although  to solve other problems,  such as monopole and domain walls \cite{Inflation}, 
 inflation with the  slow-roll conditions seems still required. 

After first developing the general formulas of linear scalar perturbations  without specifying a particular gauge and matter fields in Sec. III.B,
 we have investigated 
several possible gauge choices in Sec. III.C, and found that, unlike the case without the U(1) symmetry \cite{WM}, now various gauge choices become
possible, including the ``generalized" longitudinal gauge, synchronous gauge, and quasilongitudinal gauge.  

Applied the general formulas to a single scalar field,  in Sec. IV  we have first shown that the flat FRW universe has the same dynamics as that given in GR. As 
a result, all the results obtained in GR are also applicable here in the HMT setup, as far as only the background is concerned, including the slow roll
conditions. Then, we have found that  in the super-horizon regions,  the perturbations become adiabatic, and the comoving curvature perturbation is constant, though for a different reason from that in GR.  

In Sec. V, we have shown  
 that a master equation [cf. Eq.(\ref{6.38})]
for the scalar perturbations exists, in contrast to the case without the U(1) symmetry \cite{WM}. In addition, we have  also shown explicitly that in the sub-horizon
 regions, the metric and scalar field are tightly coupled and have the 
same oscillating frequencies.%, while in the  super-horizon regions,  the scale-invariance are the same as those given in GR.
 
We have also calculated the power spectra and spectrum indices of both the scalar and tensor perturbations  in the slow-roll approximations (Sec. VI), by using 
the uniform approximations \cite{Hab}, and expressed them explicitly in terms of the slow roll parameters and the coupling constants of high order curvature terms.
{We've found that, with some reasonable conditions on the coupling coefficients $c_{1,2}$ and $V_{i}$  [cf. the discussions presented after Eq.(\ref{6.91}) and Eqs.(\ref{6.72ba}) and (\ref{bs})], 
the spectrum index of the tensor perturbations is the same as the value given in GR, whereas the index of the scalar perturbation is a function of $\lambda$ and can be different from the standard GR value.} The power spectra are in general
different from those of GR.  For more general cases, the power spectrum $P_{\cal{R}}$  and the ratio $r$ are given in Appendix D.
We have also found that inflation in the HMT setup produces all the observational features of the universe \cite{EK}. Therefore, as far as slow-roll
inflation is concerned, the HL theory are consistent with observations. 
 
% It would be extremely interesting to find some distinguishable features of the theory from other ones, including GR, so the theory can be under direct tests. 
%%Our studies presented in this paper show that such differences can be achieved, if the gauge field $A$ and potential terms $c_1, c_2, V_n$ are 
%%not restricted to slow-evolution in time.  % roll approximations. % the background equations.  

%
%%Those include scale-invariance, non-Gaussianity \cite{IKM,Chen} and relics of 
%%  gravitational waves \cite{KDM}.
%%  

~\\{\bf Acknowledgements:}   
The work of AW was supported in part by DOE  Grant, DE-FG02-10ER41692 and NNSFC 11075141; and 
QW  was supported in part by  NNSFC grant, 11047008.

\section*{Appendix A:  Field Equations}
 \renewcommand{\theequation}{A.\arabic{equation}} \setcounter{equation}{0}

 For the action (\ref{2.4}),    the  Hamiltonian and momentum constraints are given, respectively, by,
 \bqn 
 \lb{eq1}
& & \int{ d^{3}x\sqrt{g}\left[{\cal{L}}_{K} + {\cal{L}}_{{V}} - \varphi {\cal{G}}^{ij}\nabla_{i}\nabla_{j}\varphi 
- \big(1-\lambda\big)\big(\Delta\varphi\big)^{2}\right]}\nb\\
& & ~~~~~~~~~~~~~~~~~~~~~~~~~~~~~
= 8\pi G \int d^{3}x {\sqrt{g}\, J^{t}},\\
\lb{eq2}
& & \nabla^{j}\Big[\pi_{ij} - \varphi  {\cal{G}}_{ij} - \big(1-\lambda\big)g_{ij}\nabla^{2}\varphi \Big] = 8\pi G J_{i},
 \eqn
where
 \bqn 
  \lb{eq2b}
  J^{t} &\equiv& 2 \frac{\delta\left(N{\cal{L}}_{M}\right)}{\delta N},\;\;\; J_{i} \equiv - N\frac{\delta{\cal{L}}_{M}}{\delta N^{i}}, \nb\\
   \pi^{ij} &\equiv&   \frac{\delta(N{\cal{L}}_{K})}{\delta\dot{g}_{ij}} = 
   - K^{ij} +  \lambda K g^{ij}.
 \eqn
 Variation of the action (\ref{2.4}) with respect to   $\varphi$ and $A$ yield, 
\bqn
\lb{eq4a}
& & {\cal{G}}^{ij} \Big(K_{ij} + \nabla_{i}\nabla_{j}\varphi\Big) + \big(1-\lambda\big)\Delta \Big(K + \Delta \varphi\Big) \nb\\
& & ~~~~~~~~~~~~~~~~~~~~~~~~~~~~~~~~ = 8\pi G J_{\varphi}, \\
\lb{eq4b}
& & R - 2\Lambda_{g} =   8\pi G J_{A},
\eqn
where
\bq
\lb{eq5}
J_{\varphi} \equiv - \frac{\delta{\cal{L}}_{M}}{\delta\varphi},\;\;\;
J_{A} \equiv 2 \frac{\delta\left(N{\cal{L}}_{M}\right)}{\delta{A}}.
\eq
On the other hand,  the dynamical equations now read\footnote{Note that the dynamical equations given here differ from 
those given in \cite{HW} because here we took $N^i$ as the fundamental variable instead of $N_i$ as what we did in \cite{HW}. They are both self-consistent if $N^i$ and $N_i$ are used consistently.},
 \bqn \lb{eq3}
&&
\frac{1}{N\sqrt{g}}\Bigg\{\sqrt{g}\Big[\pi^{ij} - \varphi {\cal{G}}^{ij} - \big(1-\lambda\big) g^{ij} \Delta \varphi\Big]\Bigg\}_{,t}  
\nb\\
& &~~~ = -2\left(K^{2}\right)^{ij}+2\lambda K K^{ij} - \frac{2}{N}\pi^{k(i}\nabla_k N^{j)}\nb\\
& &  ~~~~~ + \nabla_{k}\Big[\frac{N^k}{N} \pi^{ij}-(1-\lambda) F^k_{\varphi} g^{ij}\Big]\nb\\
& &  ~~~~~ - 2\big(1-\lambda\big) \Big[\big(K + \Delta \varphi\big)\nabla^{i}\nabla^{j}\varphi + K^{ij}\Delta \varphi\Big]\nb\\
& &  ~~~~~ + 2 (1-\lambda)\nabla^{(i}\left[\nabla^{j)} \varphi \left( K + \Delta\varphi\right)\right] \nb\\ 
& &  ~~~~~ +  \frac{1-\lambda}{N}\Delta\varphi \nabla^{(i}N^{j)} \nb\\
%& & ~~~~~ + \big(1-\lambda\big) \Big[2\nabla^{(i}F^{j)}_{\varphi} - g^{ij}\nabla_{k}F^{k}_{\varphi}\Big]\nb\\
& & ~~~~~ +  \frac{1}{2} \Big({\cal{L}}_{K} + {\cal{L}}_{\varphi} + {\cal{L}}_{A} + {\cal{L}}_{\lambda}\Big) g^{ij} \nb\\
& &  ~~~~~    + F^{ij} + F_{\varphi}^{ij} +  F_{A}^{ij} + 8\pi G \tau^{ij},
 \eqn
where $\left(K^{2}\right)^{ij} \equiv K^{il}K_{l}^{j},\; f_{(ij)}
\equiv \left(f_{ij} + f_{ji}\right)/2$, and
 \bqn
\lb{eq3a} 
F^{ij} &\equiv&
\frac{1}{\sqrt{g}}\frac{\delta\left(-\sqrt{g}
{\cal{L}}_{V}\right)}{\delta{g}_{ij}}
 = \sum^{8}_{s=0}{g_{s} \zeta^{n_{{\scriptscriptstyle S}}}
 \left(F_{s}\right)^{ij} },\nb\\
F_{\varphi}^{ij} &=&  \sum^{3}_{n=1}{F_{(\varphi, n)}^{ij}},\nb\\
F_{\varphi}^{i} &=&  \Big(K + \nabla^{2}\varphi\Big)\nabla^{i}\varphi + \frac{N^{i}}{N} \Delta \varphi, \nb\\
F_{A}^{ij} &=& \frac{1}{N}\left[AR^{ij} - \Big(\nabla^{i}\nabla^{j} - g^{ij}\Delta \Big)A\right],\nb\\ 
 \eqn
with   
$n_{{\scriptscriptstyle S}} =(2, 0, -2, -2, -4, -4, -4, -4,-4)$.   $ \left(F_{s}\right)_{ij}$ and $F_{(\varphi, n)}^{ij}$ are given by
  \cite{WM,WW}, 
%$F_{ij}$ and $F_{\varphi}^{ij}$ appearing in Eq.(\ref{eq3}) are given, respectively,  by \cite{WM},
 \bqn
\lb{a.1}
(F_{0})_{ij}&=& -\frac12g_{ij},\nb\\
(F_{1})_{ij}&=&-\frac12g_{ij}R+R_{ij},\nb\\
(F_{2})_{ij} &=&-\frac12g_{ij}R^2+2RR_{ij}-2\nabla_{(i}\nabla_{j)}R\nb\\
                     & & +2g_{ij}\nabla^2R,\nb\\
(F_{3})_{ij}&=&-\frac12g_{ij}R_{mn}R^{mn}+2R_{ik}R^k_j-2\nabla^k\nabla_{(i}R_{j)k}\nb\\
                     && +\nabla^2R_{ij}+g_{ij}\nabla_m\nabla_nR^{mn},\nb\\
(F_{4})_{ij}&=&-\frac12g_{ij}R^3+3R^2R_{ij}-3\nabla_{(i}\nabla_{j)}R^2\nb\\
                    && +3g_{ij}\nabla^2R^2,\nb\\
(F_{5})_{ij}&=&-\frac12g_{ij}RR^{mn}R_{mn}+R_{ij}R^{mn}R_{mn}\nb\\
                    &&+ 2RR_{ki}R^k_j  -\nabla_{(i}\nabla_{j)}\left(R^{mn}R_{mn}\right)\nb\\
                    && - 2\nabla^n\nabla_{(i}RR_{j)n}  +g_{ij}\nabla^2\left(R^{mn}R_{mn}\right)\nb\\
                    &&  + \nabla^2\left(RR_{ij}\right)   +g_{ij}\nabla_m\nabla_n\left(RR^{mn}\right),\nb\\
(F_{6})_{ij}&=&-\frac12g_{ij}R^m_nR^n_pR^p_m+3R^{mn}R_{ni}R_{mj}\nb\\
                      && +\frac32\nabla^2\left(R_{in}R^n_j\right)   + \frac32g_{ij}\nabla_k\nabla_l\left(R^k_nR^{ln}\right)\nb\\
                      & & -3\nabla_k\nabla_{(i}\left(R_{j)n}R^{nk}\right),\nb\\
(F_{7})_{ij}&=&-\frac12g_{ij}(\nabla R)^2+ \left(\nabla_iR\right)\left(\nabla_jR\right) -2R_{ij}\nabla^2R\nb\\
&& +2\nabla_{(i}\nabla_{j)}\nabla^2R-2g_{ij}\nabla^4R,\nb\\
(F_{8})_{ij}&=& -\frac12g_{ij}\left(\nabla_p R_{mn}\right)\left(\nabla^p R^{mn}\right) -\nabla^4R_{ij}\nb\\
&&  + \left(\nabla_i R_{mn}\right)\left(\nabla_j R^{mn}\right) +2\left(\nabla_p R_{in}\right)\left(\nabla^p R^n_j\right)\nb\\
&&  +2\nabla^n\nabla_{(i}\nabla^2R_{j)n}+2\nabla_n\left(R^n_m\nabla_{(i}R^m_{j)}\right)\nb\\
&& -2\nabla_n\left(R_{m(j}\nabla_{i)}R^{mn}\right)-2\nabla_n\left(R_{m(i}\nabla^nR^m_{j)}\right)\nb\\
 && -g_{ij}\nabla^n\nabla^m\nabla^2R_{mn}, 
\eqn 
and  
\bqn 
  \lb{eq3c}
F_{(\varphi, 1)}^{ij} &=& \frac{1}{2}\varphi\Big\{\Big(2K + \nabla^{2}\varphi\Big) R^{ij}  \nb\\
& & ~~~~~ - 2 \Big(2K^{j}_{k} + \nabla^{j} \nabla_{k}\varphi\Big) R^{ik}\nb\\
& & ~~~~~ - 2 \Big(2K^{i}_{k} + \nabla^{i} \nabla_{k}\varphi\Big) R^{jk}\nb\\
& &~~~~~ 
- \Big(2\Lambda_{g} - R\Big) \Big(2K^{ij} + \nabla^{i} \nabla^{j}\varphi\Big)\Big\},\nb\\
F^{ij}_{(\varphi,2)} &=& \frac{1}{2} \nabla_k\Bigg[ 2 \left(\varphi{\cal{G}}^{k(i}\nabla^{j)}\varphi \right)-\varphi{\cal{G}}^{ij} \left( \frac{2 N^k}{N} +\nabla^k\varphi\right) \Bigg] \nb\\
&& + \frac{2\varphi}{N} {\cal{G}}^{k(i} \nabla_k N^{j)}, \nb\\
%F_{(\varphi, 2)}^{ij} &=& \frac{1}{2}\nabla_{k}\left\{\varphi{\cal{G}}^{ik}  
%\Big(\frac{2N^{j}}{N} + \nabla^{j}\varphi\Big) \right. \nb\\
%& & \left.
%+ \varphi{\cal{G}}^{jk}  \Big(\frac{2N^{i}}{N} + \nabla^{i}\varphi\Big) 
%-  \varphi{\cal{G}}^{ij}  \Big(\frac{2N^{k}}{N} + \nabla^{k}\varphi\Big)\right\}, \nb\\   
F_{(\varphi, 3)}^{ij} &=& \frac{1}{2}\left\{2\nabla_{k} \nabla^{(i}f^{j) k}_{\varphi}  
- \nabla^{2}f_{\varphi}^{ij}   - \left(\nabla_{k}\nabla_{l}f^{kl}_{\varphi}\right)g^{ij}\right\},\nb\\
\eqn
where
\bqn
\lb{eq3d}
f_{\varphi}^{ij} &=& \varphi\left\{\Big(2K^{ij} + \nabla^{i}\nabla^{j}\varphi\Big) %\right.\nb\\
%& & \left. ~~~~~~~~ 
- \frac{1}{2} \Big(2K + \nabla^{2}\varphi\Big)g^{ij}\right\}.\nb\\
\eqn

 The stress 3-tensor $\tau^{ij}$ is defined as
 \bq \label{tau}
\tau^{ij} = {2\over \sqrt{g}}{\delta \left(\sqrt{g}
 {\cal{L}}_{M}\right)\over \delta{g}_{ij}}.
 \eq

The  conservation laws of energy and momentum of  matter fields read, respectively, 
 \bqn \lb{eq5a} & &
 \int d^{3}x \sqrt{g} { \left[ \dot{g}_{kl}\tau^{kl} -
 \frac{1}{\sqrt{g}}\left(\sqrt{g}J^{t}\right)_{, t}  
 +   \frac{2N_{k}}  {N\sqrt{g}}\left(\sqrt{g}J^{k}\right)_{,t}
  \right.  }   \nb\\
 & &  ~~~~~~~~~~~~~~ \left.   - 2\dot{\varphi}J_{\varphi} -  \frac{A} {N\sqrt{g}}\left(\sqrt{g}J_{A}\right)_{,t}
 \right] = 0,\\
\lb{eq5b} & & \nabla^{k}\tau_{ik} -
\frac{1}{N\sqrt{g}}\left(\sqrt{g}J_{i}\right)_{,t}  - \frac{J^{k}}{N}\left(\nabla_{k}N_{i}
- \nabla_{i}N_{k}\right)   \nb\\
& & \;\;\;\;\;\;\;\;\;\;\;- \frac{N_{i}}{N}\nabla_{k}J^{k} + J_{\varphi} \nabla_{i}\varphi - \frac{J_{A}}{2N} \nabla_{i}A
 = 0.
\eqn

\section*{Appendix B:  Scalar Fields}
\renewcommand{\theequation}{B.\arabic{equation}} \setcounter{equation}{0}

How matter couples 
with gravity in the HMT setup has not yet been worked out in the general case. In this Appendix, we shall consider the coupling of a
scalar field $\chi$ with gravity by the prescription given in \cite{Silva}. The case with detailed balance condition softly breaking was
studied in \cite{BLW}.

\subsection{ Coupling of a Scalar Field}

When the scalar field coupled only with $N,\; N^{i},\; g_{ij}$, the most general action takes the form \cite{WWM,KK},
\bq
\lb{5.1}
S_{\chi}^{(0)}\Big(N, N^{i}, g_{ij}; \chi\Big) = \int{dt d^{3}x N\sqrt{g}{\cal{L}}^{(0)}_{\chi} \Big(N, N^{i}, g_{ij}; \chi\Big)},  
\eq
where
\bqn
\lb{5.2}
{\cal{L}}^{(0)}_{\chi} &=& \frac{f(\lambda)}{2N^{2}}\Big(\dot{\chi} - N^{i}\nabla_{i}\chi\Big)^{2} - {\cal{V}},\nb\\
{\cal V}  &=& V\left(\chi\right) + \left({1\over2}+V_{1}\left(\chi\right)\right) (\nabla\chi)^2
+  V_{2}\left(\chi\right){\cal{P}}_{1}^{2}\nb\\
& & +  V_{3}\left(\chi\right){\cal{P}}_{1}^{3}  +
V_{4}\left(\chi\right){\cal{P}}_{2} + V_{5}\left(\chi\right)(\nabla\chi)^2{\cal{P}}_{2}\nb\\
& & 
+ V_{6}(\chi) {{\cal P}}_{1} {\cal{P}}_{2},
\eqn
with $V(\chi)$ and $V_{n}(\chi)$ being arbitrary functions of $\chi$, and
\bq
\lb{5.2a}
{\cal{P}}_{n} \equiv \Delta^{n}\chi.
\eq
Note that in the kinetic term we added a factor $f(\lambda)$, which is an arbitrary 
function of $\lambda$, subjected to the requirements: (i) The scalar field must be ghost-free in all the energy scales, including the UV and IR.
 (ii) In the IR limit, the scalar field has a well-defined velocity, which should be equal or very closed to its relativistic value.   (iii) The stability 
 condition in the IR requires \cite{BLW},
 \bq
 \lb{f}
 f(\lambda) > 0.
 \eq

To couple with the gauge field $A$ and the Newtonian prepotential $\varphi$, we make the replacement \cite{Silva}, 
\bq
\lb{5.3}
S_{\chi}^{(0)} \Big(N, N^{i}, g_{ij};\; \chi\Big) \rightarrow S_{\chi}\Big(N, N^{i}, g_{ij},  A, \varphi;  \; \chi\Big),
\eq
where
\bqn
\lb{5.4}
 & & S_{\chi}\Big(N, N^{i}, g_{ij},  A, \varphi; \; \chi\Big) \equiv  S_{\chi}^{ A}(\chi, A)\nb\\
 & &~~~~~~~~~~ + S_{\chi}^{(0)} \Big(N, \big(N^{i} + N\nabla^{i}\varphi\big), g_{ij}; \; \chi\Big),
 \eqn
 with 
 \bqn
 \lb{5.5}
& & S_{\chi}^{A} \equiv  \int{dt d^{3}x \sqrt{g}\Big[c_{1}(\chi)\Delta\chi + c_{2}(\chi)\big(\nabla\chi\big)^{2}\Big]} {\big(A - {\cal{A}}\big)}.\nb\\
\eqn
Thus,  the action can be cast in the form,
\bq
\lb{5.6}
S_{\chi}  =  \int{dt d^{3}x N\sqrt{g}{\cal{L}}_{\chi}}, 
\eq
where
\bqn
\lb{5.7}
{\cal{L}}_{\chi} &=& {\cal{L}}^{(0)}_{\chi} + {\cal{L}}^{(A,\varphi)}_{\chi},\nb\\
{\cal{L}}^{(A,\varphi)}_{\chi} &=& \frac{A - {\cal{A}}}{N}  \Big[c_{1}\Delta\chi + c_{2}\big(\nabla\chi\big)^{2}\Big]\nb\\
&&  - \frac{f}{N}\Big(\dot{\chi} - N^{i}\nabla_{i}\chi\Big)\big(\nabla^{k}\varphi\big)\big( \nabla_{k}\chi\big)\nb\\
& & + \frac{f}{2}\Big[\big(\nabla^{k}\varphi\big)\big(\nabla_{k}\chi\big)\Big]^{2},
 \eqn
with $ {\cal{L}}^{(0)}_{\chi}$  given by Eq.(\ref{5.2}). Then, we find that  
\bqn
\lb{5.8a}
J^t &=&  -2\Bigg(\frac{f}{2N^2}\Big(\dot{\chi}-N^k\nabla_k \chi\Big)^2+{\cal V}\Bigg) \nb\\
& & -\Big[c_1 \triangle \chi+c_2 \big(\nabla\chi\big)^{2}\Big]\big(\nabla\varphi\big)^{2}\nb\\
& & + f \Big[\big(\nabla^k\varphi\big)\big(\nabla_k\chi\big)\Big]^2,  \\
\lb{5.8b}
J^i &=& \frac{f}{N}\Bigg[\dot{\chi}-\Big(N^k + N \nabla^{k}\varphi\Big)\big(\nabla_k \chi\big)\Bigg] \nabla^i \chi \nb\\
      & &   + \Big[c_1 \triangle \chi+c_2 \big(\nabla\chi\big)^{2}\Big]\nabla^{i}\varphi,\\
\lb{5.8c} 
J_\varphi &=&\frac{1}{N\sqrt{g}}\Bigg\{\sqrt{g}\Big[c_1 \triangle \chi+c_2 \big(\nabla\chi\big)^{2}\Big]\Bigg\}_{,t}   \nb\\
&&-\frac{1}{N}\nabla_i\Bigg\{f\Big[\dot{\chi} - \big(N^{k} + N\nabla^{k}\varphi\big)\big(\nabla_{k}\chi\big)\Big]\nabla^{i}\chi\nb\\
& & ~~~~~~~~~ +  \Big[c_1 \triangle \chi+c_2 \big(\nabla\chi\big)^{2}\Big]\nb\\
& & ~~~~~~~~~~~ \times \Big(N^i+N\nabla^i\varphi\Big)\Bigg\}, \\
\lb{5.8d}
J_A &=& 2\Big[c_1 \triangle \chi+c_2 \big(\nabla\chi\big) ^{2}\Big],\\
\lb{5.8e}
\tau_{ij} &=& \tau_{ij}^{(0)} +\tau_{ij}^\varphi, 
\eqn
where
\bqn
\lb{5.9}
 \tau^{(0)}_{ij}   &=&   g_{ij} \Big\{{\cal{L}}^{(0)}_{\chi} + \nabla_{k}\big[\big({\cal{V}}_{,1} +\Delta{\cal{V}}_{,2}\big)\nabla^{k}\chi  + {\cal{V}}_{,2}\nabla^{k}\Delta\chi\big]\Big\}\nb\\
 & & +  \big(1+ 2V_1+2V_5 {\cal P}_2\big)  \left(\nabla_{i} \chi\right) \left(\nabla_{j} \chi\right) \nb\\
& & - 2 \big(\nabla_{(i}{\cal{V}}_{,1}\big)\big(\nabla_{j)}\chi\big)- 2 \big(\nabla_{(i}\Delta{\cal{V}}_{,2}\big)\big(\nabla_{j)}\chi\big)\nb\\
& & - 2 \big(\nabla_{(i}{\cal{V}}_{,2}\big)\big(\nabla_{j)}\Delta\chi\big),\nb\\
\tau_{ij}^\varphi &=&  g_{ij}\Big\{{\cal{L}}^{(A,\varphi)}_{\chi} - \frac{1}{N}\nabla_{k}\big[c_{1}\big(A - {\cal{A}}\big)\nabla^{k}\chi\big]\Big\}\nb\\
& & + \frac{2({\cal{A}} - A)}{N}\Big[c_1\nabla_{i}\nabla_{j}\chi + c_2 \big(\nabla_{i}\chi\big)\big(\nabla_{j}\chi\big)\Big]\nb\\
& & + \Big[c_1\Delta\chi + c_2 \big(\nabla\chi\big)^{2}\Big]  \big(\nabla_{i}\varphi\big)\big(\nabla_{j}\varphi\big)\nb\\
& & + \frac{2f}{N}\Big[\dot{\chi} - \big(N^{k} + N\nabla^{k}\varphi\big)\big(\nabla_{k}\chi\big)\Big]\Big(\nabla_{(i}\chi\Big)\Big(\nabla_{j)}\varphi\Big)\nb\\
& & + \frac{2}{N}\nabla_{(i}\Big[c_{1}\big(A - {\cal{A}}\big)\nabla_{j)}\chi\Big],
  \eqn
 and
   \bqn
 \lb{5.9a}
{\cal V}_{,1} &\equiv& \frac{\partial {\cal V}}{\partial
{\cal{P}}_{1}}
    =  2V_{2} {\cal{P}}_{1} + 3 V_{3} {\cal{P}}_{1}^{2}
   + V_{6} {\cal{P}}_{2},\nb\\
{\cal V}_{,2} &\equiv& \frac{\partial {\cal V}}{\partial
{\cal{P}}_{2}}
    =   V_{4} + V_{5} (\nabla\chi)^2 + V_{6} {\cal{P}}_{1}.
 \eqn
 
 On the other hand, the variation of the action (\ref{5.6}) with respect to $\chi$ yields the following generalized Klein-Gordon equation,
  \bqn
   \lb{5.10}
   & & \frac{f}{N\sqrt{g}} \Bigg\{\frac{\sqrt{g}}{N}\Big[\dot{
\chi} - \left(N^{k}+N \nabla^{k} \varphi\right)\nabla_{k}\chi\Big]\Bigg\}_{,t}  \nb\\
&&  = \frac{f}{N^{2}} \nabla_{i}\Bigg\{\Big[\dot{ \chi} -\left( N^{k}+ N \nabla^{k}\varphi \right) \nabla_{k} \chi\Big] \left(N^{i}+ N \nabla^{i}\varphi \right)\Bigg\}\nb\\
 && ~~~  + \frac{2g^{ij}}{N} \nabla_{(i}\Bigg\{\nabla_{j)}\Big[(A-{\cal{A}})c_{1}\Big]-(A-{\cal{A}})c_{2} \nabla_{j)}\chi\Bigg\} \nb\\ 
 & & ~~~ + \frac{A-{\cal{A}}}{N} \Big[c_{1}'\Delta\chi + c_{2}'\big(\nabla\chi\big)^{2}\Big]\nb\\
  & & ~~~ +  \nabla^{i}\Big[\big( 1+2V_1+2V_5 {\cal P}_2 \big)\nabla_{i} \chi\Big] \nb\\
 && ~~~  -  {\cal V}_{, \chi} - \Delta\left({\cal V}_{,1}\right) - \Delta^{2}\left({\cal V}_{,2}\right),
  \eqn
where $c_{1}' \equiv dc_{1}(\chi)/d\chi$, and
  \bqn
 \lb{5.11}
{\cal V}_{, \chi} &\equiv& \frac{\partial {\cal V}}{\partial
\chi} =  V'  + V_{1}' (\nabla\chi)^2
   +  V_{2}'{\cal{P}}_{1}^{2}  +  V_{3}'{\cal{P}}_{1}^{3}  \nb\\
  & &   + V_{4}'{\cal{P}}_{2} + V_{5}'
  (\nabla\chi)^2{\cal{P}}_{2} + V'_{6}{\cal{P}}_{1}{\cal{P}}_{2}.
 \eqn

 \subsection{Linear Perturbations under the Newtonian Quasilongitudinal Gauge}

Under  the gauge (\ref{NLG}),   Eqs.(\ref{5.17a}) - (\ref{5.19}) can be cast in the forms, 
  \bqn
 \lb{6.17a}
 &&\int d^{3}x\Bigg\{\partial^2\psi
- \frac{1}{2}\big(3\lambda -1\big){\cal H}
\Big[3\psi' +\partial^2B\Big]\Bigg\} \nb\\ 
&& =4\pi G\int d^{3}x\Bigg\{f\bar{\chi}' \delta{\chi}' + \Big(a^2{V}' + \frac{V_{4}}{a^{2}} \partial^4\Big) \delta \chi\Bigg\}, ~~~~~\\
 \lb{6.17b}
 && \int d^3 x a^2 \bar{\chi}' \Bigg\{f \delta \chi'' + 2 {\cal H} f \delta \chi' + a^2 {V}'' \delta \chi -3 f \bar{\chi}' \psi' \nb\\
 && ~~~~~~~~~~~~~~~~~ - \frac{\bar{A}}{\bar{\chi}'}\frac{\left(a c_1 \delta \chi\right)'}{a}\Bigg\}\nb\\
 &&= -\int d^3 x \; \partial^4\Bigg\{{V}_{4}  \delta \chi' + \Big({V}'_{4} \bar{\chi}' - {V}_{4} {\cal H}\Big)  \delta \chi\Bigg\},\\
 \lb{6.17c}
 && (3\lambda -1){\psi}'    - (1- \lambda)\partial^2 B = 8\pi G f \bar{\chi}' \delta \chi, \\
  \lb{6.17d}
& & 2{\cal{H}}\psi +  \big(1-\lambda\big)\big(3\psi'   + \partial^2B\big) 
\nb\\
&&   ~~~~~ = 8\pi G\Big[\big({c}'_{1} \bar{\chi}'+ {c}_{1}{\cal H} - f \bar{\chi}'\big) \delta \chi + {c}_1  \delta \chi'\Big] ,\\
  \lb{6.17e}
&& \psi = 4\pi G {c}_1  \delta \chi,  \\
 \lb{6.17f}
 &&  \psi'' + 2{\cal{H}}\psi'   + \frac{1}{3}\partial^{2}\big(B' + 2{\cal{H}}B\big)\nb\\ 
& & ~~~~
 - \frac{2}{3(3\lambda-1)}\Bigg(1 + \frac{\alpha_{1}}{a^{2}}\partial^{2}
   + \frac{\alpha_{2}}{a^{4}}\partial^{4}\Bigg)\partial^{2}\psi \nb\\
   & & ~~~~ + \frac{2}{3(3\lambda - 1)a}\partial^{2}\big(\hat{A}\psi - \delta{A} \big)\nb\\ 
%   & & ~~~~ + \frac{\lambda - 1}{(3\lambda - 1)a}\partial^{2}\big(\delta\varphi' +  {\cal{H}}\delta\varphi\big)  \nb\\
   & & ~~~~  = \frac{8\pi G }{3\lambda-1}\Big(f\bar{\chi}' \delta \chi' - a^2{V}'  \delta \chi\Big), \\
   \lb{6.17g}
 &&  \psi - \big(B' + 2{\cal{H}}B\big)  %- \phi   
+ \frac{1}{a^{2}}\Big(\alpha_{1} + \frac{\alpha_{2}}{a^{2}}\partial^{2}\Big)\partial^2\psi\nb\\
& & ~~~ -\frac{1}{a}\Big(\hat{A}\psi  - \delta{A} \Big) = 0,  \\
 \lb{6.17h}
 && f \Big\{\delta \chi'' + 2 {\cal H} \delta \chi' - \bar{\chi}' \left[3 \psi' +\partial^2 B\right]\Big\} + a^2V''\delta\chi\nb\\
 &&   ~~~~ = 2 \Bigg(\frac{1}{2}+{V}_{1} -\frac{{V}_{2}+{V}'_{4}}{a^2}\partial^2 -\frac{{V}_{6}}{a^4} \partial^4\Bigg)\partial^2 \delta \chi\nb\\
 && ~~~~~~~~~~ +\frac{1}{a} \partial^2\Big[2\hat{A}\left({c}'_{1} - {c}_{2}\right)  \delta \chi  + c_1 \delta{A}\Big].
 \eqn
Recall  $\hat{A}=a\bar{A}$. It can be shown that Eqs.(\ref{6.17d}) and (\ref{6.17f}) are not independent, and can be 
obtained from the others. Therefore, in the present case there are four independent differential equations,
(\ref{6.17c}), (\ref{6.17e}), (\ref{6.17g}), and (\ref{6.17h}),  for the four unknowns, $\psi,\; B, \; \delta{A}$ and $\delta\chi$.

\section*{Appendix C:  Quantization of scalar perturbations}
\renewcommand{\theequation}{C.\arabic{equation}} \setcounter{equation}{0}

 This part summarizes the discussions given in \cite{Inflation}.\footnote{Note that here we did not consider any modifications of the commutation relations
  \cite{AHM}.} To quantize the scalar field, suppose we have the normalized action of second order
 \bq
 \lb{b.01}
 S^{(2)} = \frac{1}{2} \int d \eta d^3 x \Big[v'^2 - \beta(\eta, \partial^n) v^2\Big],
 \eq
 where  $\beta(\eta, \partial^n)$ is in general time-dependent explicitly and $\partial^n = \partial^2,  \partial^4,  \partial^6 , ...$. Now promote the field $v$ and its conjugate momentum to operators,
 \bq
 \lb{b.02}
 v \rightarrow \hat{v}(\eta, \mathbf{x}) = \int \frac{d^3k}{(2\pi)^3} \Big[v_k(\eta) \hat{a}_{\mathbf{k}}e^{i \mathbf{k}\mathbf{x}} + v^{*}_k(\eta) \hat{a}^{\dagger}_{\mathbf{k}}e^{-i \mathbf{k}\mathbf{x}} \Big].
 \eq
 If we define the Fourier image of $v(\eta, \mathbf{x})$ as
 \bq
 \lb{b.03}
 v(\tau, \mathbf{x}) = \int \frac{d^3k}{(2\pi)^3} v_{\mathbf{k}} (\eta) e^{i \mathbf{k}\mathbf{x}},
 \eq
 this is equivalent to say that
 \bq
 \lb{b.04}
 v_{\mathbf{k}} (\eta) \rightarrow \hat{v}_{\mathbf{k}} (\eta) = v_k(\eta) \hat{a}_{\mathbf{k}} + v^{*}_k(\eta) \hat{a}^{\dagger}_{\mathbf{k}}.
 \eq
 $v_k(\tau)$ are called the \textit{mode functions}. They satisfy the second order classical equation of motion (EoM)
 \bq
 \lb{b.05}
 v''_k + \beta(\eta, k^{2n})v_k =0.
 \eq
 The canonical commutation relation between quantum field $\hat{v}_\mathbf{k}$ and its conjugate momentum $\hat{v}'_\mathbf{k}$ is,
 \bq
 \lb{b.06}
 \bra{0}[\hat{v}_\mathbf{k}, \hat{v}'_\mathbf{k}]\ket{0}= i\hbar.
 \eq
 If we want to have 
 \bq
 \lb{b.07}
 [\hat{a}_{\mathbf{k}}, \hat{a}^{\dagger}_{\mathbf{k}'}]=(2\pi)^3 \delta^3(\mathbf{k}-\mathbf{k}'),
 \eq
 the norm (Wronskian) has to be
 \bq
 \lb{b.08}
 v^*_k v'_k - v^{*'}_k v_k = -i \hbar.
 \eq
 Besides the normalization condition, we need another boundary condition to determine the mode functions completely. 
 Usually this is obtained by requiring that the vacuum state to be the ground state of the Hamiltonian back in the far past when the mode is deep inside horizon
 \bq
 \lb{b.09}
 \hat{H} \ket{0} = E_0\ket{0},
 \eq
 where the vacuum is defined as $\hat{a}_{\mathbf{k}} \ket{0} =0$. Since we have
 \bq
 \lb{b.10}
 \hat{H} = \frac{1}{2} \left( \hat{v}'^2_\mathbf{k} + \beta \hat{v}^2_\mathbf{k}\right), 
 \eq
 this requires $v'_k = \pm i \sqrt{\beta} v_k$ if $ \ket{0}$ is the ground state. Thus
 \bq
 \lb{b.11}
 v_k = C e^{-i \int d\eta\sqrt{\beta}},
 \eq
 where the positive frequency branch is selected to ensure the positivity of normalization, and $C$ will be determined by the normalization condition (\ref{b.08}).
 
 \section*{Appendix D:  General expressions for power spectra}
\renewcommand{\theequation}{D.\arabic{equation}} \setcounter{equation}{0}
Here we give the more general expressions of spectra without setting $b_1=b_2=b_3=1=\beta_0$. $M_{\text{A}}=M_{\text{B}} = M_{*}$ (and thus by (\ref{6.82}) $g^2_3=g_8$) and $\bar{A} =0$ is still assumed. In the limiting case when $\epsilon_{{\scriptscriptstyle \text{HL}}} \equiv (H/M_*)^2 \ll 1$, we find
\bqn
\lb{C1.1}
P_{{\cal{R}}} (k) &\simeq& {P^{\text{GR}}_{{\cal{R}}}} \frac{1}{(b_1)^{2/3}\beta_0} \left(1+ \frac{c_1 \dot{\bar{\chi}}}{2 M^2_{\text{pl}}H}\right)^2 \nb\\
&& \times \Bigg[1-\frac{9 b_2}{4 b^2_1} \epsilon_{{\scriptscriptstyle \text{HL}}}+ \frac{81(17 b^2_2 - 8 b_1b_3)}{128b^4_1}\epsilon^2_{{\scriptscriptstyle \text{HL}}}\Bigg], \nb\\
\\
\lb{C1.2}
r &\simeq& 16\epsilon_{{\scriptscriptstyle V}}  (b_1)^{2/3}\beta_0 \left(1+ \frac{c_1 \dot{\bar{\chi}}}{2 M^2_{\text{pl}}H}\right)^{-2}\nb\\
&& \times \Bigg[1+\frac{9 (b_2-2 b^2_1)}{4 b^2_1} \epsilon_{{\scriptscriptstyle \text{HL}}}\nb\\
&& ~~~~~~ + \frac{81(36b^4_1-17 b^2_2 + 8 b_1b_3)}{128b^4_1}\epsilon^2_{{\scriptscriptstyle \text{HL}}}\Bigg] .
\eqn

In the limit $\epsilon_{{\scriptscriptstyle \text{HL}}} \gg 1$, we obtain
\bqn
\lb{C2.1}
P_{{\cal{R}}} (k) &\simeq& {P^{\text{GR}}_{{\cal{R}}}}  \frac{4e^{\frac{1}{2}} \eta_{{\scriptscriptstyle \text{HL}}}}{9\sqrt{b_3}\beta_0} \left(1+ \frac{c_1 \dot{\bar{\chi}}}{2 M^2_{\text{pl}}H}\right)^2\nb\\
&& \times \Bigg[1-\frac{b_2}{2b_3}\left(\frac{4b_3}{9}\eta_{{\scriptscriptstyle \text{HL}}}\right)^{\frac{1}{3}}\Bigg], \\
\lb{C2.2}
r &\simeq& 16\epsilon_{{\scriptscriptstyle V}} \frac{\sqrt{b_3} \beta_0}{2} \left(1+ \frac{c_1 \dot{\bar{\chi}}}{2 M^2_{\text{pl}}H}\right)^{-2} \nb\\
&& \times \Bigg\{1 -\left[\frac{1}{4}-\frac{b_2}{2 b_3}\left(\frac{b_3}{4}\right)^{1/3}\right]\left(\frac{16}{9}\eta_{{\scriptscriptstyle \text{HL}}}\right)^{\frac{1}{3}}\Bigg\} . \nb\\
\eqn
  Clearly, the magnitude of the ratio $r$ are dependent on the values of $b_1$ and $b_3$.

\end{document}